\documentclass[twocolumn]{aastex631} 
\usepackage[version=3]{mhchem}
\usepackage{comment}

\newcommand{\mum}{$\rm \mu \mathrm{m}~$}

\newcommand{\OIII}{[\textsc{Oiii}]\,}
\newcommand{\OII}{[\textsc{Oii}]\,}
\newcommand{\HII}{H{\sc ii}~}

\usepackage{color}

\newcommand{\reply}[1]{\textcolor{black}{#1}}
\newcommand{\replytwo}[1]{\textcolor{black}{#1}}

\shorttitle{\OIII Luminosity calculation in First Light}
\shortauthors{Nakazato et al.}

\graphicspath{{./}{figures/}}
\newcommand{\figdir}{./} 

\begin{document}

\title{Simulations of high-redshift [\textsc{OIII}]
emitters: Chemical evolution
and multi-line diagnostics}
\correspondingauthor{Yurina Nakazato}
\email{yurina.nakazato@phys.s.u-tokyo.ac.jp}
\author[0000-0002-0984-7713]{Yurina Nakazato}
\affiliation{Department of Physics, The University of Tokyo, 7-3-1 Hongo, Bunkyo, Tokyo 113-0033, Japan}

\author[0000-0001-7925-238X]{Naoki Yoshida}
\affiliation{Department of Physics, The University of Tokyo, 7-3-1 Hongo, Bunkyo, Tokyo 113-0033, Japan}
\affiliation{Kavli Institute for the Physics and Mathematics of the Universe (WPI), UT Institute for Advanced Study, The University of Tokyo, Kashiwa, Chiba 277-8583, Japan}
\affiliation{Research Center for the Early Universe, School of Science, The University of Tokyo, 7-3-1 Hongo, Bunkyo, Tokyo 113-0033, Japan}

\author[0000-0002-8680-248X]{Daniel Ceverino}
\affiliation{Universidad Autonoma de Madrid, Ciudad Universitaria de Cantoblanco, E-28049 Madrid, Spain}
\affiliation{CIAFF, Facultad de Ciencias, Universidad Autonoma de Madrid, E-28049 Madrid, Spain}

\begin{abstract}
Recent observations by James Webb Space Telescope discovered a number of high-redshift galaxies with strong emission lines from doubly ionized oxygen. Combined with ALMA observations of far-infrared lines, multi-line diagnostics can be applied to the high-redshift galaxies in order to probe the physical conditions of the inter-stellar medium.
We study the formation and evolution of galaxies using the FirstLight simulation suite, which provides outputs of 62 high-resolution, zoom-in galaxy simulations. We devise a physical model of \textsc{Hii} regions and calculate spatially resolved \OIII line emission. 
We show that massive galaxies with stellar masses of $M_* > 10^9 M_\odot$ chemically evolve rapidly to $z=9$. Young stellar populations in the star-forming galaxies boost the \OIII line emission, rendering the ratio of line luminosity to star formation rate larger than that for low-redshift galaxies, which is consistent with recent observations. 
Measuring the flux ratios of rest-frame optical and far-infrared lines allows us to estimate the physical conditions such as density and metallicity of the star-forming gas in high-redshift \OIII emitters.
\end{abstract}

\keywords{}

\section{Introduction} \label{sec:intro}
Understanding the formation and evolution of the first galaxies is one of the key scientific goals of new generation telescopes
including James Webb Space Telescope (JWST) and Atacama Large Millimetre/Submillimetre Array (ALMA). High-redshift galaxies can be detected and identified using strong emission lines,  
among which \OIII 88\mum line is 
thought to be promising \citep{Inoue:2014}.
A number of galaxies have been found at 
$z>7$ by ALMA observations targeting the \OIII 88$\mu\mathrm{m}$ line \citep[e.g.][]{Inoue:2016, Hashimoto:2018}, 
including the most distant galaxy candidate at $z = 13.27$ with a $4\sigma$ \OIII 88$\mu\mathrm{m}$ detection \citep{Harikane:2022_HDrop}. Since the \OIII line emission originates from \textsc{Hii} regions around young massive stars, it can be used to trace the star formation activities and also the physical properties of the inter-stellar medium (ISM) in the early galaxies.

JWST is opening a new window into the early universe through its superb observational capability in near-infrared.
For example, JWST Early Research Observation (ERO) in the lensing field SMACS 0723 already reported three galaxies confirmed spectroscopically by NIRSpec \citep{Schaerer:2022, Curti:2022, Heintz:2022}. NIRSpec instrument is capable of detecting and identifying various rest-frame optical lines such as \OII 3727\AA, \OIII 4959\AA\, and \OIII 5007\AA. 
The relatively weak \OIII 4363\AA\ line has been detected for all the three galaxies, enabling us to estimate the ISM metallicity in a direct manner.

Detailed numerical simulations are indispensable to study the physical conditions of the ISM. There have been several studies focusing on \OIII emission lines from high-z galaxies \citep{Hirschmann:2017, Olsen:2017, Moriwaki:2018, Katz:2019, Arata:2020, Ceverino:2021, Pallottini:2022}.   \citet{Moriwaki:2018} use a cosmological simulation with a large boxsize of 50 Mpc \citep{Shimizu:2016} to calculate the \OIII88$\mu\mathrm{m}$ line intensities for a few hundred galaxies with stellar masses of $\sim 10^8~M_\odot$. 
High-resolution, zoom-in simulations have
also been performed to study the internal
structure of early galaxies 
\citep{Katz:2019,Arata:2020}. 
For the upcoming observations conducted by JWST, it is urgently needed to study the population of high-redshift galaxies with high resolution in a fully cosmological context.
In this {\it Letter}, we use the outputs of FirstLight simulation \citep{Ceverino:2017}. The simulation suite is motivated to produce a statistically significant number of galaxies with very high resolution at the epoch of reionization. Thanks to the mass and volume complete sample of more than 60 massive galaxies and to the high-resolution of $\sim 20$ pc, we can investigate the internal structure as well as statistics of the high-redshift galaxies. 

Throughout this {\it Letter}, we assume $Z_\odot = 0.02$ as the solar metallicity \citep{Andres:1989}. 

\section{method} \label{sec:method}
\subsection{Cosmological Simulation} \label{subsec:simulation}
We use mass-limited galaxy samples selected from the FirstLight simulation suite \citep{Ceverino:2017}. The simulations are performed with ART code \citep{Kravtsov:1997, Kravtsov:2003, Ceverino:2009, Ceverino:2014}, which follows gravitational $N$-body dynamics and Eulerian hydrodynamics using an adaptive mesh refinement method. Besides the two processes, the code incorporates astrophysical processes relevant for galaxy formation. The so-called subgrid physics includes atomic and molecular cooling of hydrogen and helium, photoionization heating by a cosmological UV background with partial self-shielding, and star formation and the associated stellar feedback. Details are described in \citet{Ceverino:2017}. The simulations track metals released from SNe-Ia and from SNe-II, using supernovae yields from \citet{Woosley:1995}. 

Our simulated galaxies are hosted by dark matter haloes with maximum circular velocity ($V_\mathrm{max}$) higher than 178 km/s at $z = 5$ in a cosmological volume of 40 $h^{-1}$Mpc on a side.
The host haloes are selected in a low-resolution $N$-body only simulation, for which refined initial conditions are generated using a standard zoom-in technique \citep{Klypin:2011}. The refinement achieves the dark matter particle mass of $m_\mathrm{DM} = 8\times 10^4~M_\odot$, the minimum star particle mass of $10^3 ~M_\odot$, and the maximum spatial resolution  
is a few tens proper parsec depending on the
refinement level.

We calculate the stellar mass distribution for the selected 62 massive galaxies at $z=9, 8, 7, 6$. The maximum stellar mass is 9.5, 9.7, 10.1, 10.7$\times 10^9~M_\odot$, respectively. The sample allows us to study the evolution of more massive galaxies than in previous simulations, e.g., \citet{Moriwaki:2018}, SERRA simulation \citep{Pallottini:2022} and S\'{I}GAME simulation \citep{Olsen:2017}, and thus is well-suited to compare with observed massive galaxies by HST, ALMA, and JWST \citep[e.g.][]{Tacchella:2022, Graziani:2020, Topping:2022, Trussler:2022, Barrufet:2022, Leethochawaliit:2022}. 



\subsection{Line emissivity calculation}\label{subsec:line_emissivity_calculation}
We generate emission-line maps for our galaxy samples by choosing a region enclosed by 0.3 times the virial radius of the host halo as same as \citet{Mandelker:2014, Mandelker:2017}. We configure a uniform 3D grid with a side length of 100 pc.We locate the star particles and gas elements within each grid, and store the mass of stars younger than 10 Myr, the average density of the gas with $n_\mathrm{H} > 0.1~\mathrm{cm^{-3}}$, 
and the average metallicity of the cold/warm gas with $T < 5\times 10^4~\mathrm{K}$. 
These physical quantities assigned to the individual grids are then used to compute 
the line emissivities in a similar manner to those in \citet{Hirschmann:2017, Moriwaki:2018, Ceverino:2021}. 




\replytwo{The luminosities of emission lines from \HII regions are calculated as follows:}
\begin{align}
    L_\mathrm{line} &= (1- f_\mathrm{esc})\,C_\mathrm{line}(Z_\mathrm{gas}, U, n_\textsc{Hii})\,L^{\mathrm{caseB}}_\mathrm{H\beta}, \label{eq:L_line} \\
    L^{\mathrm{caseB}}_\mathrm{H\beta} & = 4\pi j_{\mathrm{H\beta}} V=h \nu_{\mathrm{H\beta}}\left(\frac{\alpha_{\mathrm{H\beta}}^{\mathrm{eff}}}{\alpha_{\mathrm{B}}}\right) Q, \label{eq:L_Hbeta}
\end{align}
where $f_\mathrm{esc}$ is the Lyman continuum escape fraction and $C_\mathrm{line}$ is \replytwo{line emissivities relative to ${\rm H}\beta$, which are calculated by CLOUDY \citep{Ferland:2013} in the same manner as in \citet{Inoue:2014} and \citet{Moriwaki:2018}. We calculate the H$\beta$ line luminosity with the case-B approximation \citep{Dopita:2003} in eq.(\ref{eq:L_Hbeta}).}

The $\mathrm{H\beta}$ emission rate per unit volume per unit time per unit solid angle 
is denoted as $j_{\mathrm{H\beta}}$, and $\alpha_{\mathrm{H\beta}}^{\mathrm{eff}}$ is an effective recombination coefficient, $Q$ is the production rate of ionizing photons from each star particle, and $\alpha_\mathrm{B}$ is the case-B hydrogen recombination coefficient given by
\begin{equation}
    \alpha_\mathrm{B} = 2.6\times 10^{-13}\left(\frac{T_\mathrm{e}}{10^4~\mathrm{K}}\right)^{-0.85} ~\mathrm{cm^3 s^{-1}}
\end{equation}
with a constant electron temperature $T_\mathrm{e} = 10^4~\mathrm{K}$.

\reply{It has been suggested that 
galaxies at high redshift may have 
large escape fractions of 
$f_\mathrm{esc} \sim 0.1-0.4$, both 
by radiative transfer simulations for massive galaxies with $M_\mathrm{halo} > 10^{10-11}M_\odot$ \citep{Yajima:2011, Kimm_Cen:2014, Wise:2014, Paardekooper:2015, Xu:2016}
and by recent observations \citep[e.g.][]{Nakajima:2020}.
We first test a simplest model with
a constant escape fraction of $f_{\rm esc} = 0.1$.
In order to incorporate the possible scatter and variation of
$f_\mathrm{esc}$, 
we randomly assign a value in the range of [0, 0.4] using a Gaussian distribution with $\sigma = 0.1, \mu = 0.1$. By comparing the results, we find that the line luminosities $(\propto (1-f_{\rm esc}))$ change only by a factor of up to $\sim 1.5$. We thus adopt the Gaussian assignment as our fiducial model, and show the results in the following sections.}



Since individual \textsc{Hii} regions are not resolved in our simulations, we resort to a physical model of the ISM structure to calculate the line emissivities of \textsc{Hii}
regions. We characterize the ISM by the local gas density $n$ and metallicity $Z$, and also by a volume-averaged ionization parameter
\begin{equation}
    \langle U \rangle = \frac{3\alpha_\mathrm{B}^{2/3}}{4c}\left(\frac{3Q\epsilon^{2}
    n_\textsc{Hii}}{4\pi}\right)^{1/3} \label{eq:U}.
\end{equation}

Our fiducial model assumes a constant gas density $n_\textsc{Hii}$ in a spherical \textsc{Hii} region surrounding a star particle \citep[see, e.g.][]{Panuzzo:2003, Gutkin:2016}.
We set the \textsc{Hii} region density $n_\textsc{Hii} = 100 ~\mathrm{cm}^{-3}$ \citep[e.g.][]{Osterbrock:2006, Hirschmann:2017, Hirschmann:2022}. 
We define the volume-filling factor of the gas as
\begin{equation}
    \epsilon = \frac{n_\mathrm{gas, grid}}{n_{\textsc{Hii}}},
\end{equation}
where $n_\mathrm{gas, grid}$ is the gas number density in each grid. In rare cases where the volume-averaged gas density exceeds $n_\textsc{Hii}$ ($\epsilon > 1$), we set the filling factor to unity. Note that a larger $n_\mathrm{gas, grid}$ for a fixed $n_\textsc{Hii}$ yields a larger filling factor $\epsilon$.
Hence the resulting line emissivity depends only weakly on the assumed $n_\textsc{Hii}$ in our model.
We have tested with a few variations with $n_\textsc{Hii} = 50, ~ 300 ~\mathrm{cm}^{-3}$, and explicitly checked that our main findings in the following sections are not sensitively
affected by this choice. 
\replytwo{We look for the line ratio table $C_{\mathrm{line}}$ with ionization parameter $U$ and gas metallicity $Z_{\rm gas}$, which are closest to those of the target stellar particle in our simulation.}
\begin{table}[htbp]
\centering
\begin{tabular}{lr}
\hline
\vspace{0.4mm}
$\log_{10}~ (Z_\mathrm{gas}/Z_\odot)$ & -1.30, -0.70, -0.40, 0., 0.30   \\
$\log_{10}~ U$           & -4.0, -3.9, ..., -1.1, -1.0 \\
$\log_{10}~ (n_\textsc{Hii}/\mathrm{cm}^{-3}$) &  2.0 (fixed) \\
\hline
\end{tabular}
\caption{The parameters used to calculate the line luminosities with CLOUDY.}\label{table:cloudy}
\end{table}

We compute the production rate of ionizing photons $Q$ of the simulated galaxies using publicly available tables from the Binary Population and Spectral Synthesis (BPASS) model \citep{Byrne:2022}. 
Our simulations adopt a stellar initial mass function represented 
by broken power laws as 
\begin{align}
\begin{aligned}
    & N(M < M_\mathrm{max}) \propto \\
    & \int^{M_1}_{0.1}\left(\frac{M}{M_\odot}\right)^{\alpha_1} dM + M_1^{\alpha_1}\int^{M_\mathrm{max}}_{M_1}\left(\frac{M}{M_\odot}\right)^{\alpha_2} dM
\end{aligned}
\end{align}
with $\alpha_1 = -1.3, ~ \alpha_2 = -2.35, ~M_1 = 0.5,~M_\mathrm{max} = 300~ M_\odot $ as in  \citet{Ceverino:2019}. 
We use a grid of 13 values of metallicity, from $Z = 10^{-5}$ to $0.04$
, and 50 logarithmic bins in stellar population ages between 1 Myr and 100 Gyr.

 \begin{figure*}[htbp]
    \begin{tabular}{cc}
      \begin{minipage}[t]{0.49\linewidth}
        \includegraphics[keepaspectratio, scale=0.7]{\figdir/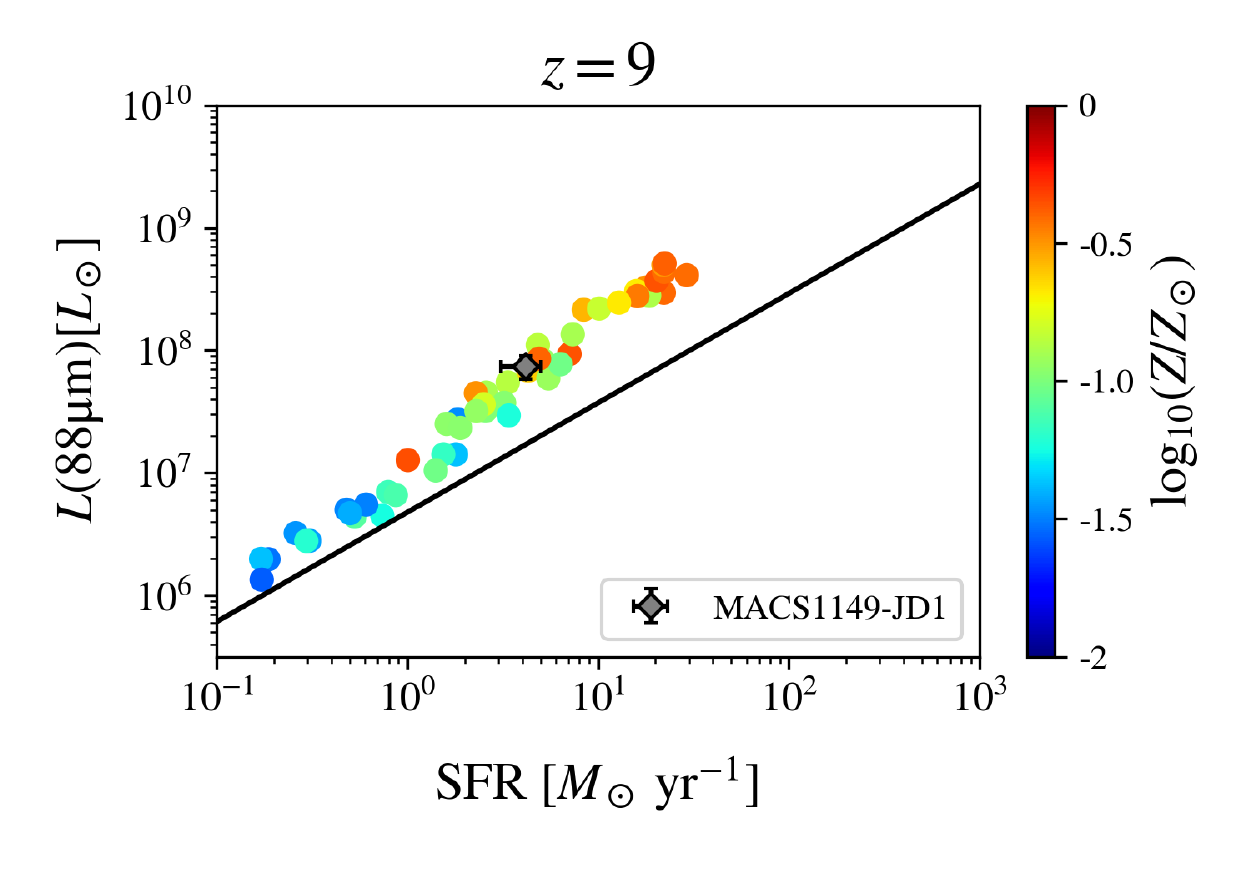}
      \end{minipage} &
      \begin{minipage}[t]{0.49\linewidth}
        \includegraphics[keepaspectratio, scale=0.7]{\figdir/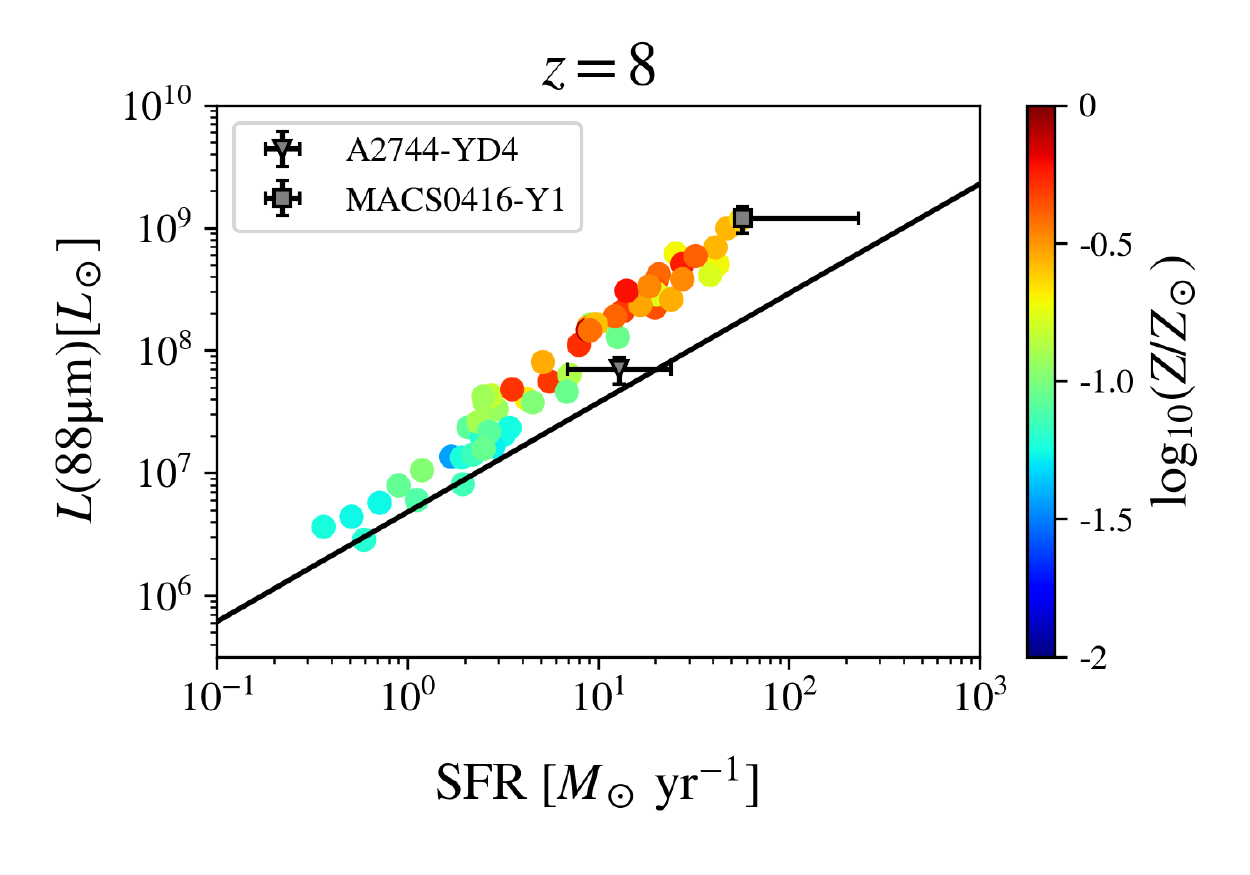}
      \end{minipage} \\
  
      \begin{minipage}[t]{0.49\linewidth}
        \includegraphics[keepaspectratio, scale=0.7]{\figdir/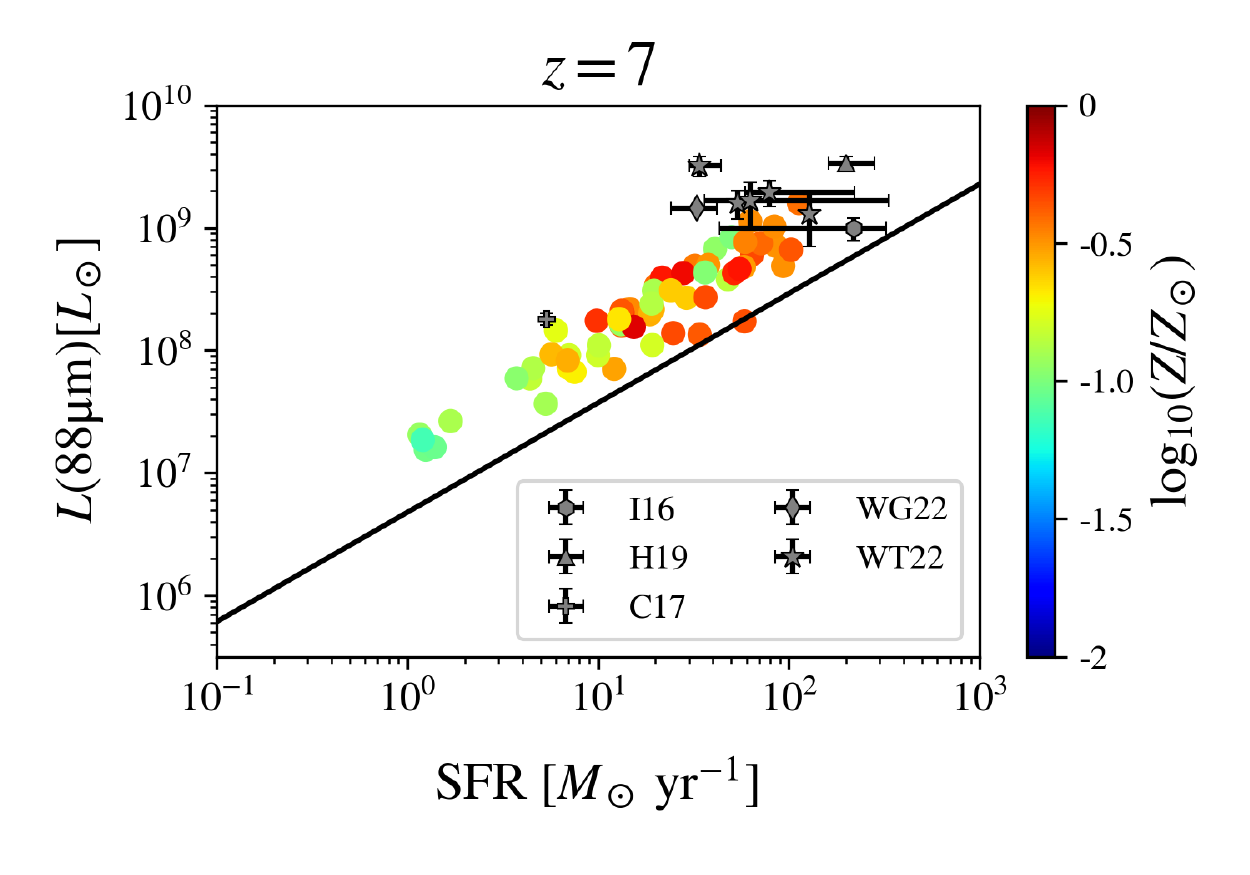}
      \end{minipage} &
      \begin{minipage}[t]{0.49\linewidth}
        \includegraphics[keepaspectratio, scale=0.7]{\figdir/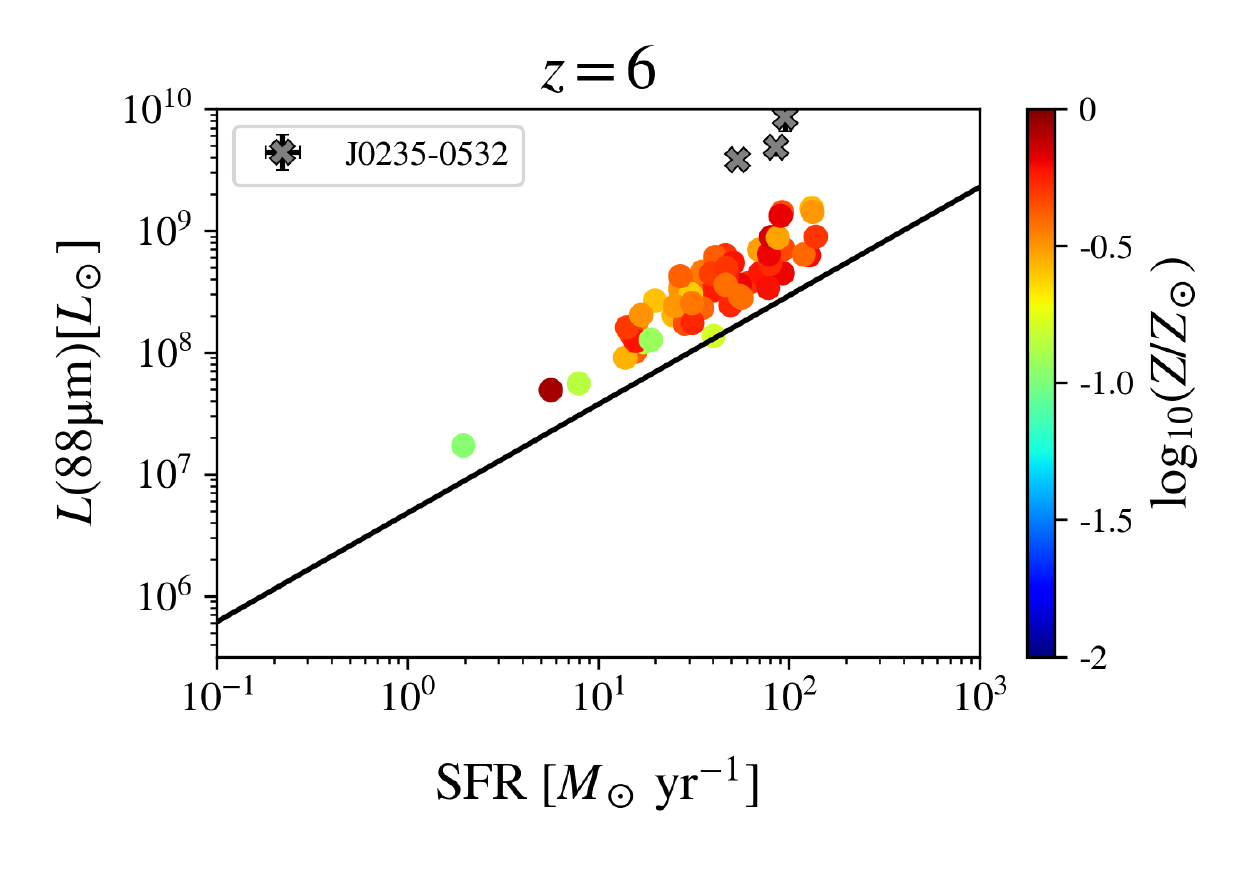}
      \end{minipage} 
    \end{tabular}
     \caption{The \OIII 88 $\mu\mathrm{m}$ luminosity versus SFR for our 62 simulated galaxies at $z = 9$ (top left), $z = 8$ (top right), $z = 7$ (bottom left), and $z=6$ (bottom right). The solid circles are colored with the gas metallicity (see the colorbar on the right). For comparison, we show the \OIII-SFR relation derived from observations of local galaxies by \citet{De_Looze:2014}. Gray points are the observational results of high-z $(z > 6)$ galaxies  from \citet{Hashimoto:2018,Laporte:2017, Tamura:2019}, \citet{Inoue:2016}(I16), \citet{Hashimoto:2019}(H19), \citet{Carniani:2017}(C17), \citet{Wong:2022}(WG22), \citet{Witstok:2022}(WT22) and \citet{Harikane:2020}. }
    \label{fig:OIII_SFR_obs}
  \end{figure*}

We re-assign "fine" ages to star particles in order to mitigate the discreteness effect caused by
our simulation set up.
Our simulations produce new star particles with a fixed time step of $\Delta t_\mathrm{SF} = 5~\mathrm{Myr}$, and the simulation output timings are not synchronized with $\Delta t_\mathrm{SF}$. 
In a snapshot, young stars typically have discretized ages such like $t_{\rm age}=2$ Myr, 7 Myr, etc. 
The apparently minor gap in stellar ages causes a large impact when we compute the line emissivities
because the ionization photon production rate quickly decreases with age.
For instance, in the BPASS SED of a single stellar population that we use, the number of ionizing photons decreases over a factor of 100 from age 1 Myr to 10 Myr \citep{Xiao:2018}.
We thus re-assign the stellar age as follows. We consider star particles younger than 15 Myr, with stamped ages at $T_1, ~T_2,~T_3~(T_1 < T_2 < T_3)$ Myr. We do random sampling within each age interval. For instance,
to a star with $T_1$, we randomly draw a new age within  $[1, T_1]$ Myr and re-assign to it. Finally, we select star particles younger than 10 Myr for our
emission line calculation.
\reply{We note that, in our CLOUDY calculation of Equation (1), we also use directly the BPASS SED so that the hydrogen ionizing photon production rate $Q$ is
calculated consistently in our model.}

We consider stellar atmosphere models with different elemental compositions, i.e., different values of $[\alpha/ \mathrm{Fe}]$. In the BPASS\_v2.3 \citep{Byrne:2022}, there are five models with the mass fractions in $\alpha$-elements relative to iron of $\Delta (\log(\alpha/\mathrm{Fe})) = -0.2, +0.0, +0.2, +0.4$ and +0.6. For the calculation of [$\alpha$/Fe], the $\alpha$-element abundance is approximated by the oxygen abundance ($\log N_\mathrm{O}$) assuming that a half of the mass in metals produced by SNII are in the form of oxygen atoms;
\begin{equation}
    \log N_\mathrm{O} = \log (f_\mathrm{O} z_\mathrm{SNII} / A_\mathrm{O}), \label{eq:N_O}
\end{equation}
where $f_\mathrm{O}, \, z_\mathrm{SNII}$ are the fraction of oxygen released by Type-II SNe, and the mass fraction of metals released from Type-II SNe, respectively. Here, the atomic weight of oxygen is $A_\mathrm{O} = 16$ and we assume $f_\mathrm{O} = 0.5$ \citep{Woosley:1995}.
We calculate the iron abundance ratio considering both contributions from Type-Ia and II SNe as
\begin{equation}
    N_\mathrm{Fe} = \frac{(f_\mathrm{Fe, Ia} \,z_\mathrm{SNIa} + f_\mathrm{Fe, II} \, z_\mathrm{SNIa})}{A_\mathrm{Fe}},
\end{equation}
where $z_\mathrm{SNIa}$ is the mass fraction of metals released from Type-Ia SNe and $A_\mathrm{Fe} = 56$. We set the fractions $f_\mathrm{Fe, Ia} = 0.5$ \citep{Thielemann:1986} and $f_\mathrm{Fe, II} = (0.026, 0.033)$ for metal mass ratio between zero and solar metallicity \citep{Nomoto:2006, Ceverino:2019}, respectively. 
Finally, [$\alpha$/Fe] is obtained from
\begin{equation}
    [\alpha/\mathrm{Fe}] = \log N_\mathrm{O} - \log N_\mathrm{Fe} - \log (N_\mathrm{O}/N_\mathrm{Fe})_\odot,
\end{equation}
where $(N_\mathrm{O}/N_\mathrm{Fe})_\odot = 1.17$ is the solar value of O/Fe abundance ratio.

\reply{
We consider ISM dust attenuation for the rest-frame optical emission line, 
[O$_{\rm III}$] 5007 \AA.
We use the results of attenuation at 1500\AA \, from \citet{Mushtaq:2023}, who conduct three-dimensional radiation transfer calculations for FirstLight galaxy samples. We adopt the approximate relation of the attenuation and the galaxy stellar mass in \citet{Mushtaq:2023}.
Assuming Calzetti law \citep{Calzetti:2000}, we use $A_{1500}/A_{5007}=2.5$ and derive 
$A_{5007} = 0.7, 0.24, 0.008$ for
galaxies with 
$\log(M_\star/M_\odot) = 10, 9, 8$, respectively.   
}

\section{results} \label{sec:results}
We focus on rest-frame sub-millimeter and optical \OIII lines from high-redshift galaxies, which are detected by ALMA and JWST.

\subsection{$L_{\OIII}$ vs SFR} \label{subsec:LOIII-SFR}
Figure \ref{fig:OIII_SFR_obs} shows the \OIII 88$\mu\mathrm{m}$ 
luminosity against star formation rate (SFR) for our galaxy samples. The color-bar indicates the nebular metallicity $Z_\mathrm{neb}$, which is the line luminosity-weighted gas metallicity. We compare with the observed local galaxies \citep{De_Looze:2014} 
and with the observed \OIII 88$\mu\mathrm{m}$ luminosities of high-redshift galaxies (see the caption).
At $z=9$ to $z = 7$, most of our simulated galaxies are located above the local galaxy relation (solid line), similar to the results of \citet{Moriwaki:2018, Arata:2020, Pallottini:2022}.

At $z = 7-9$, our galaxy samples
are distributed around the observed galaxies. It is interesting that luminous galaxies are already chemically enriched with $\log(Z/Z_\odot)\sim -0.5$ at the early epochs. 
\reply{Our simulation predicts the following relation at $z = 7-9$:
\begin{equation}
 L_{{\rm OIII},{88}} \propto \left(\frac{\mathrm{SFR}}{M_\odot\,{\rm yr}^{-1}}\right)^{0.9-1.2},
\end{equation}
which is slightly steeper than the local relationship for the entire local galaxy population \citep[Table 3 of ][]{De_Looze:2014}:
\begin{equation}
L_{{\rm OIII},{88}} \propto \left(\frac{\mathrm{SFR}}{M_\odot\,{\rm yr}^{-1}}\right)^{0.89}.
\end{equation}
}

\begin{figure*}[htbp]
\begin{center}
 \includegraphics[scale=0.7]{\figdir/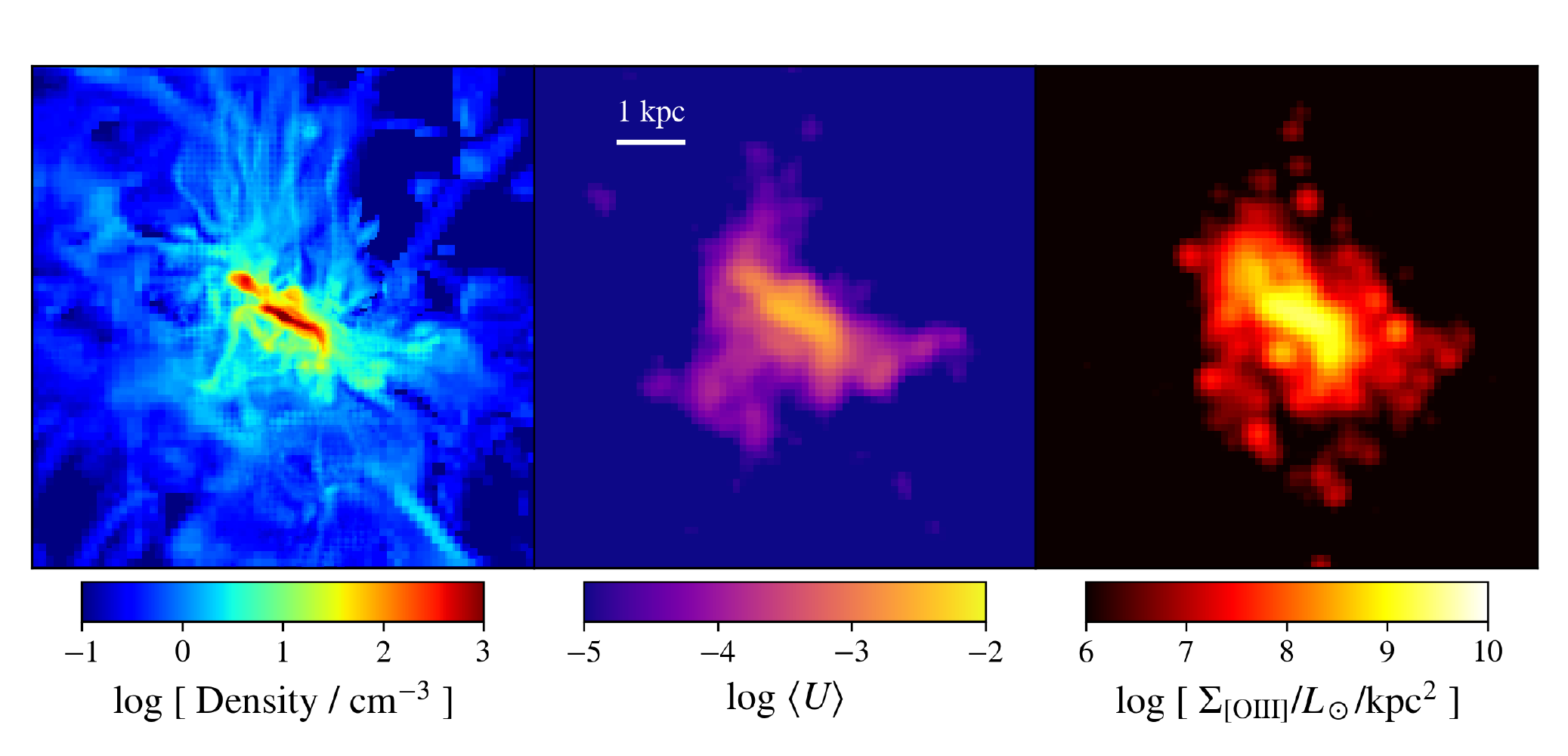}
 \caption{Projected gas density (left), averaged ionization parameter (middle), and \OIII 88\mum distribution (right) for a galaxy sample FL964 at $z =7$. Each panel shows a region with a side length and depth of 0.3$R_\mathrm{vir}(=7.4~\mathrm{kpc})$.}
   \label{fig:FL964_OIII_compare}
\end{center}
\end{figure*}

We find three galaxies with $L_{\OIII} > 10^9 ~L_\odot$ at $z = 7$, which are as bright as several observed galaxies. We study the structure of one of them (sample FL964) in detail.
It has $M_\mathrm{gas} = 6.41\times 10^9~M_\odot, ~ M_\mathrm{\star} = 9.96\times 10^9~M_\odot$, and a specific SFR of $ 11~\mathrm{Gyr}$ at $z = 7$. 
Figure \ref{fig:FL964_OIII_compare} shows the
projected maps of number density of gas, ionization parameter, and \OIII88$\mu$m. 
Clearly, regions with high ionization parameters
of $\log U \sim -2$ cause high emissivities, 
consistent with the observation by \citet{Harikane:2020}, also with recent simulations \citep{Kohandel:2022} and photoionization models \citep{Sugahara:2022}.
The total luminosity of \OIII 5007\AA\, of FL964 is 
$7.60 \times 10^9 L_\odot$, which is about 5 times larger than $L_{\OIII, 88}$.  

\subsection{The mass-metallicity relation}\label{subsec:MZR}
\begin{figure}[htbp]
\begin{center}
 \includegraphics[scale=0.7]{\figdir/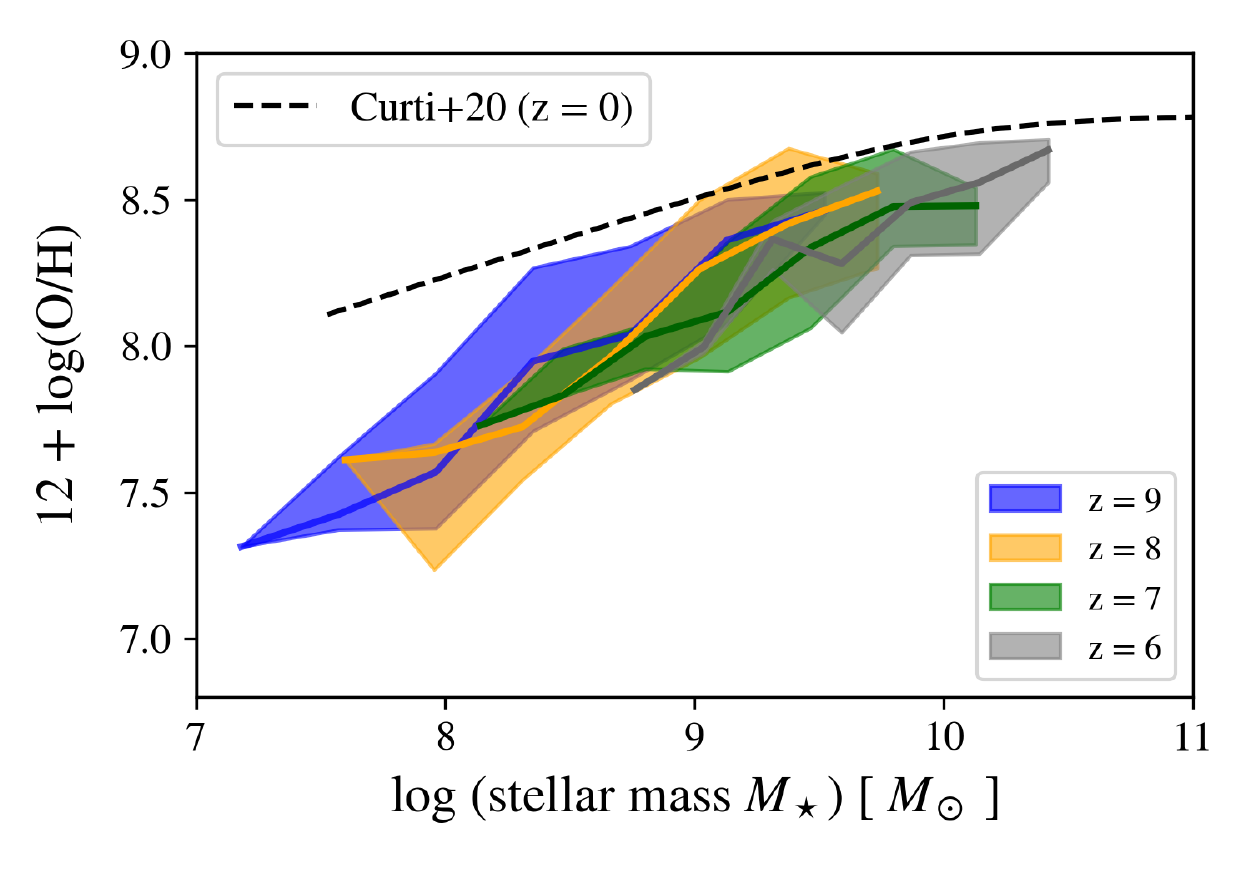}
 \caption{Gas-phase metallicity versus stellar mass for our galaxy samples from $z = 9$ to $z = 6$. The solid lines show the median and the colored bands indicate the sample dispersion in the range of 5-95\%. The dashed line is the local mass-metallicity relation from \citet{Curti:2020}. 
 }
   \label{fig:MZR}
\end{center}
\end{figure}

It is important to examine the metallicity evolution 
of our simulated galaxies. We study the so-called
mass-metallicity relation (MZR) by calculating the gas-phase metallicity for individual galaxies. 
 Figure \ref{fig:MZR} shows the stellar mass-gas phase oxygen abundance relation. 
 \reply{For each grid, we convert the metals produced in SNII, $z_{{\rm SNII}}$, into} oxygen abundance by adopting the conversion equation of \citet{Mandelker:2014};
\begin{equation}
    \frac{\mathrm{O}}{\mathrm{H}} = \frac{f_\mathrm{O}{z_{{\rm SNII}}}}{XA_\mathrm{O}}.
\end{equation}
We set the hydrogen mass fraction $X = 0.755$ and the other values of $f_\mathrm{O}$ and $A_\mathrm{O}$ are the same as those of eq.(\ref{eq:N_O}), which adopts the solar oxygen abundance $12 + \log(\mathrm{O/H}) = 8.9$. 
\reply{We first calculate the oxygen abundance for each gas element (grid), and then calculate the abundance for the entire galaxy by taking the average weighted by [O$_{\rm III}$] line emissity.}
This weighting is compatible with observational methods such as direct method or strong line method, which use oxygen emission lines
\citep[e.g.][]{Bian:2018, Izotov:2019}.

We calculate the mass of stars within the region of 0.3 $R_\mathrm{vir}$.
 \,In Figure \ref{fig:MZR}, we also plot the MZR for local galaxies from \citet{Curti:2020} (dashed line) and recent JWST observation results of high-redshift galaxies \citep{Sun:2022b, Curti:2022, Langeroodi:2022, Williams:2022}.
\citet{Curti:2022} estimated metallicities of SMACS field galaxies by direct method, \citet{Sun:2022b} adopt strong line calibration by \citet{Bian:2018} using O32, and \citet{Langeroodi:2022} and \citet{Williams:2022} adopt strong line method by \citet{Izotov:2019}.

 Our simulated galaxies have similar metallicities (oxygen abundance) and stellar masses to the observed ones. Note that Figure \ref{fig:MZR} shows the {\it evolution} for a fixed sample of simulated galaxies, rather than for all the galaxies at respective epochs. Namely, we select the galaxies at $z=5$ by mass and plot their progenitors at $z=6-9$. Hence we likely miss low-mass, low-metallicity galaxies at $z=9$ (see \citet{Langan:2020} for the mass-metallicity of low-mass galaxies in FirstLight).
 Some galaxies with $M_\star > 10^9~M_\odot$ have gas-phase metallicities of $12 + \log\mathrm{(O/H)}\sim 8.5$ even at $z=9$, suggesting that metal-enrichment can proceed rapidly in the early galaxies. 
 
 \begin{figure*}[htbp]
    \begin{tabular}{cc}
      \begin{minipage}[t]{0.49\linewidth}
        \includegraphics[keepaspectratio, scale=0.7]{\figdir/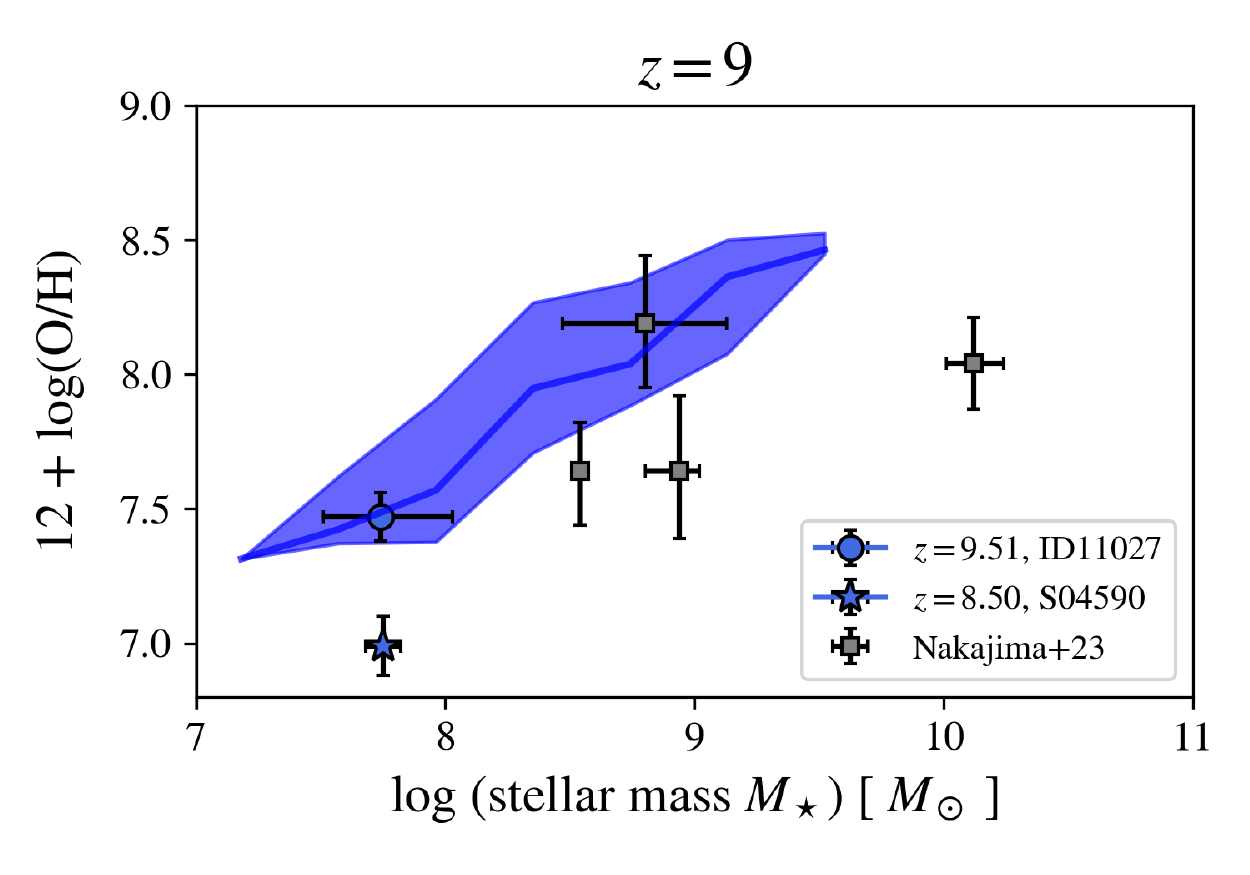}
      \end{minipage} &
      \begin{minipage}[t]{0.49\linewidth}
        \includegraphics[keepaspectratio, scale=0.7]{\figdir/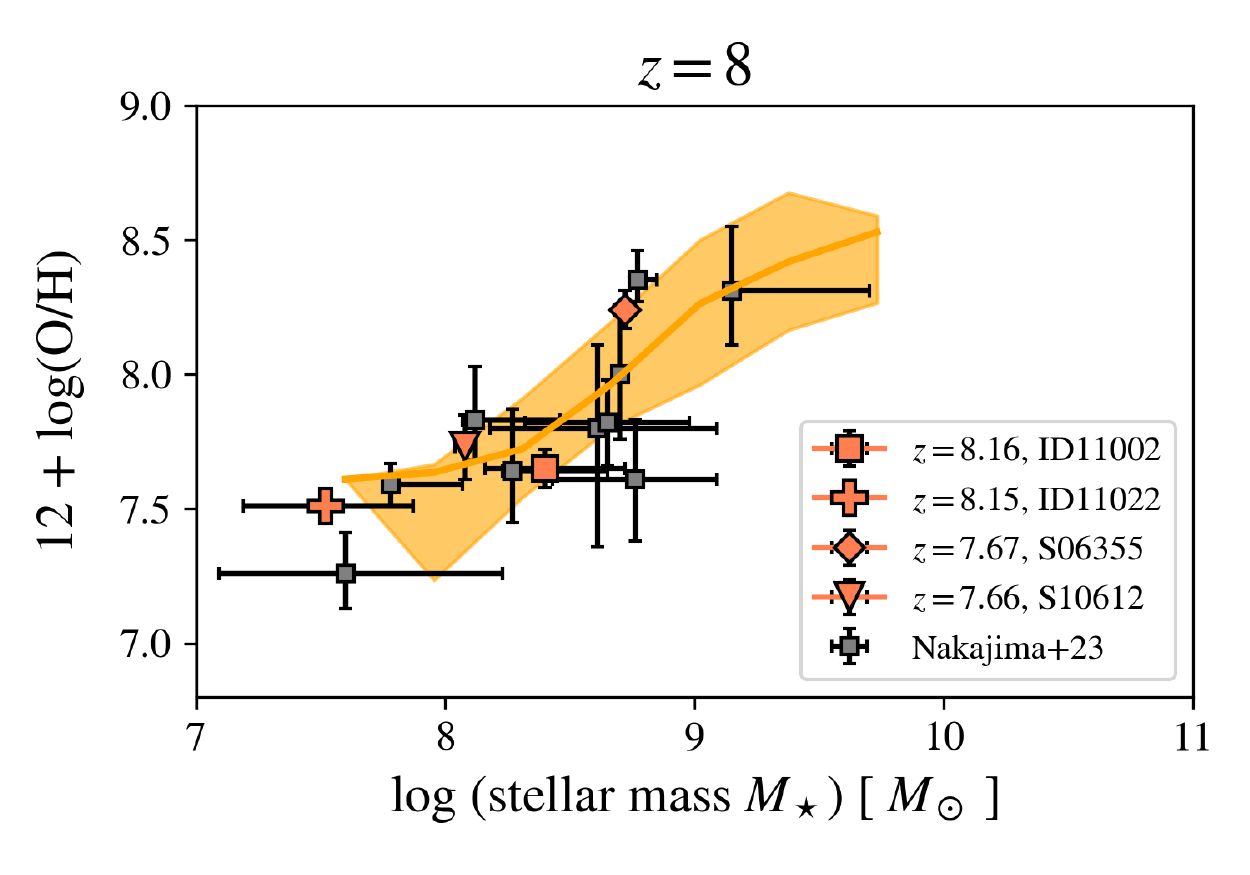}
      \end{minipage} \\
  
      \begin{minipage}[t]{0.49\linewidth}
        \includegraphics[keepaspectratio, scale=0.7]{\figdir/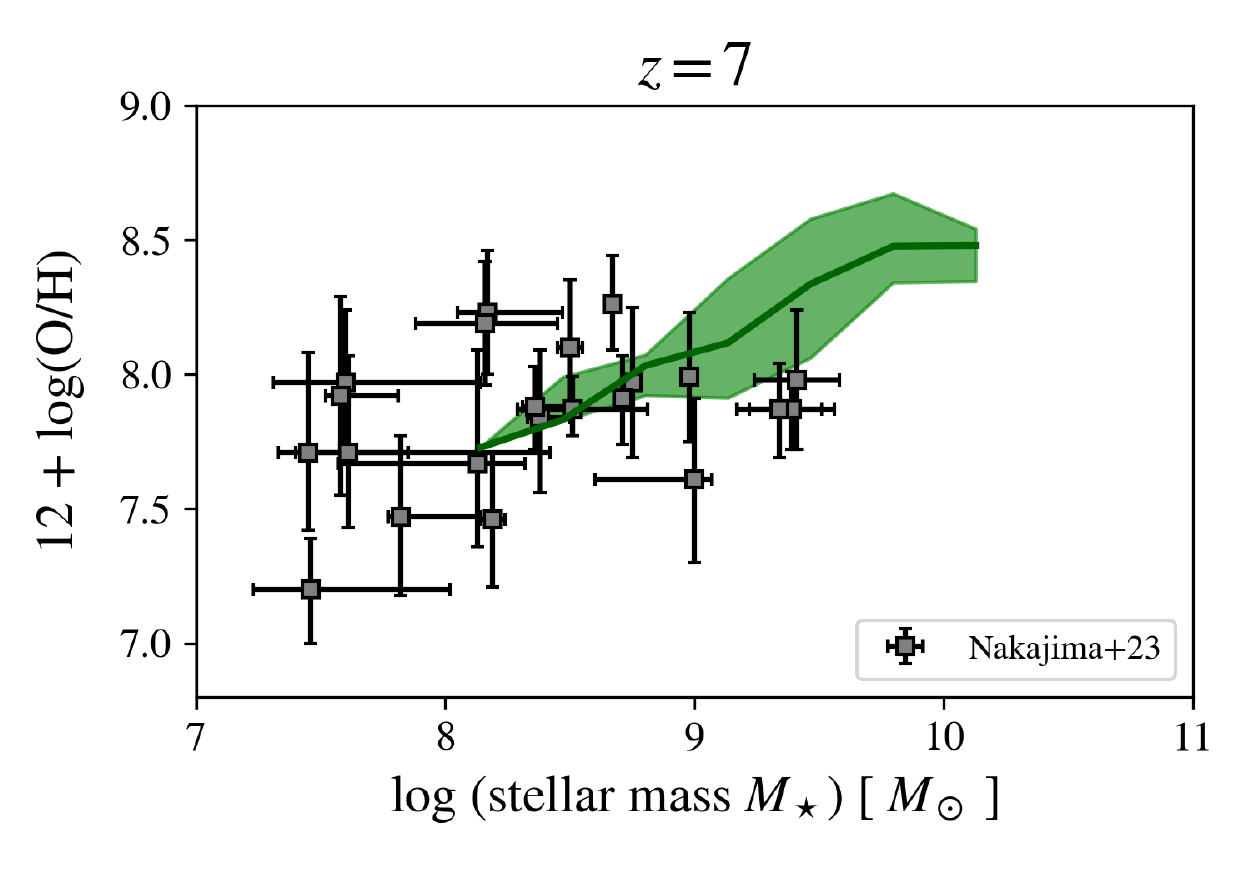}
      \end{minipage} &
      \begin{minipage}[t]{0.49\linewidth}
        \includegraphics[keepaspectratio, scale=0.7]{\figdir/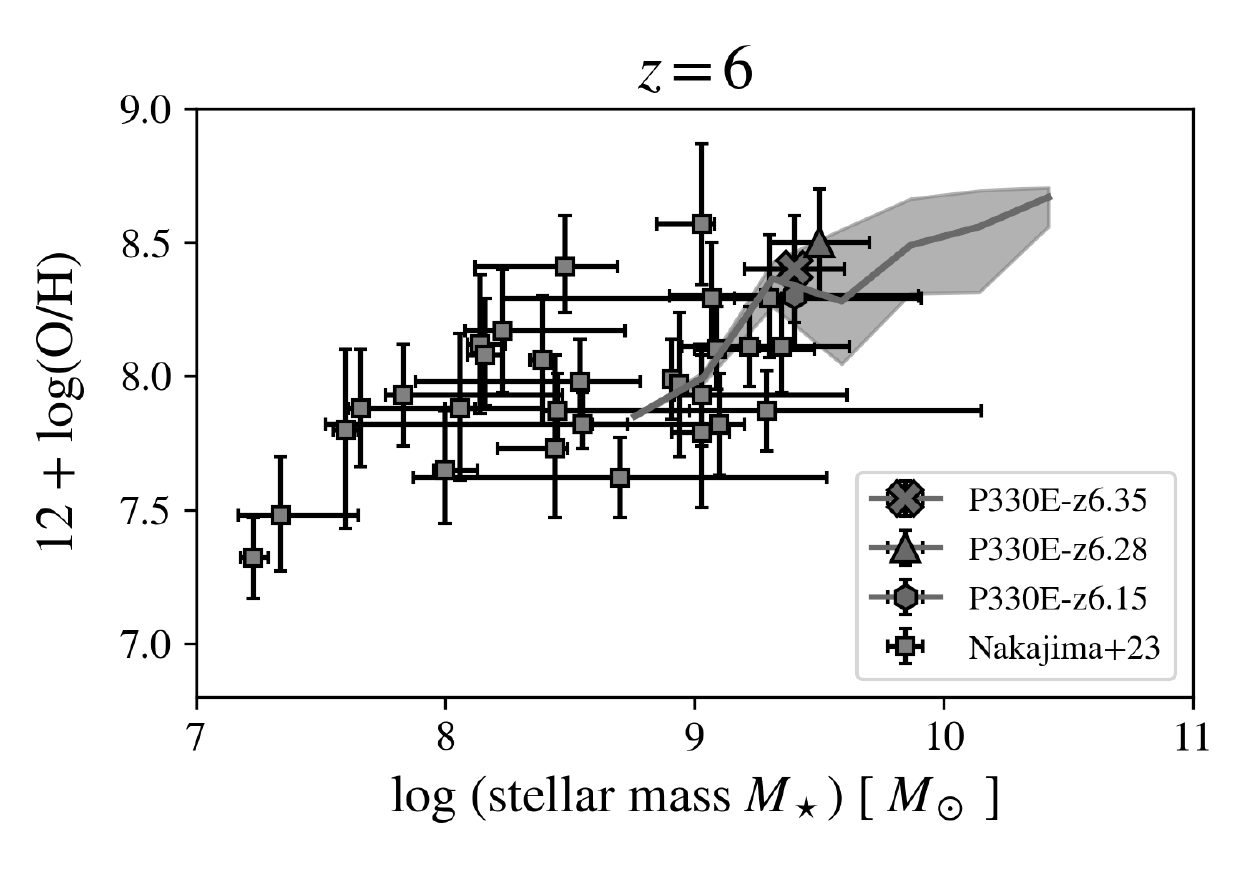}
      \end{minipage} 
    \end{tabular}
     \caption{Comparison with observed galaxies by JWST at each redshift $z=9, 8, 7,$ and 6. We plot $z > 7$ galaxies observed in SMACS J0723 field \citep{Curti:2022},  $z\sim 6$ galaxies observed by JWST/ NIRCam WFSS mode \citep{Sun:2022b}, and $z=8.1-9.5$ galaxies observed in the cluster RX J2129.4+0009 field (two galaxies at $z\sim 8.15$ from \citet{Langeroodi:2022} and one at $z=9.51$ from \citet{Williams:2022}), respectively. The gray squared plots are results from \citet{Nakajima:2023}.}
    \label{fig:MZR_obs}
  \end{figure*}
  
\subsection{Far-IR/optical line ratios}
\begin{figure*}[htbp]
\begin{center}
\includegraphics[scale=0.7, clip]{\figdir/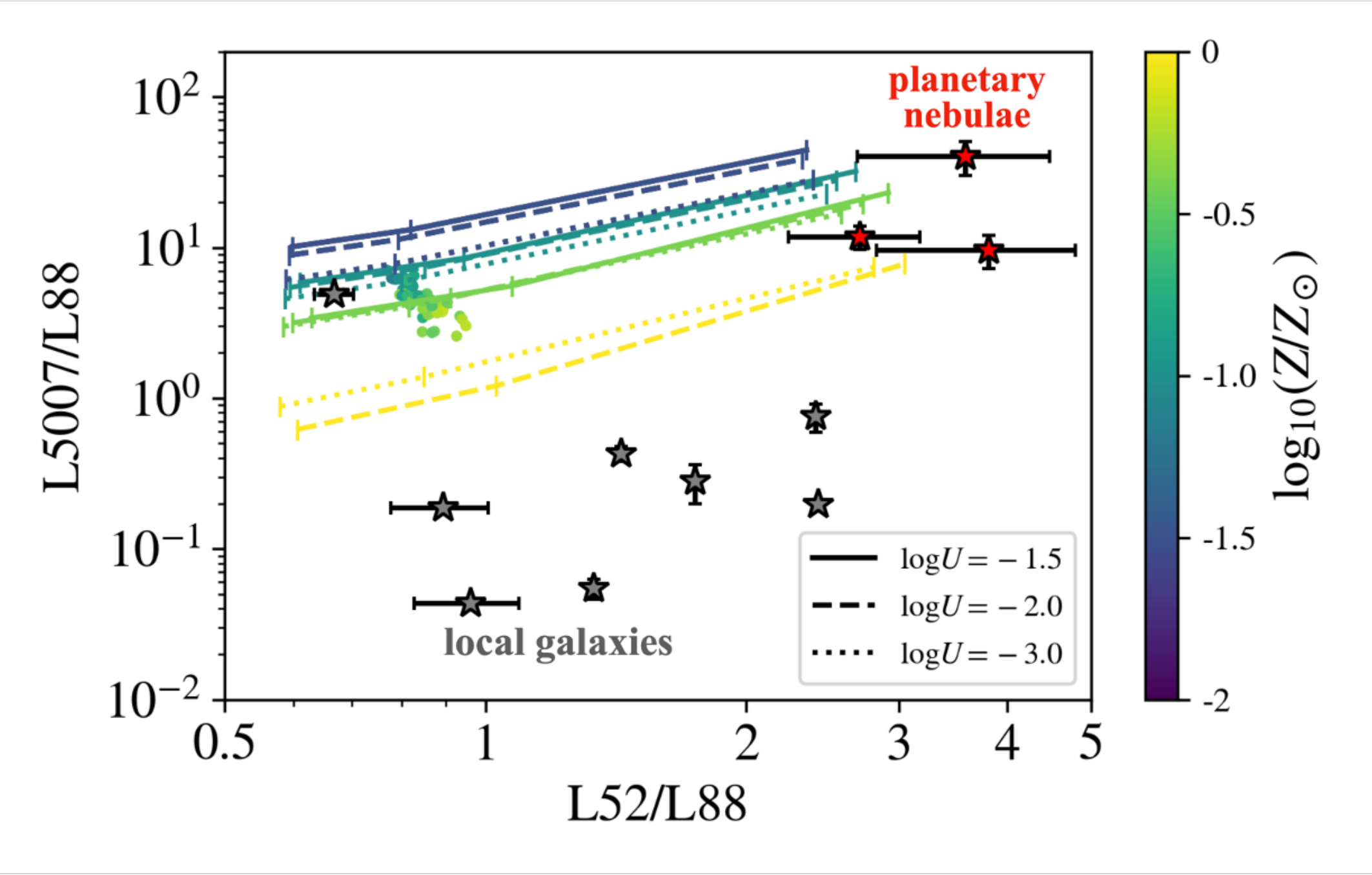}
 \caption{Line luminosity ratio $R_{5007/52}$ against $R_{52/88}$. 
 Our simulated galaxies at $z=7$ are represented by solid circles colored with gas metallicity. Gray star
 symbols show the local galaxies from \citet{Brauher:2008, Moustakas:2006} and red ones show the local planetary nebulae from \citet{Dinerstein_Lester:1985}. The results of CLOUDY calculations are represented by lines colored with metallicity ($\log (Z/Z_\odot) = -1.30, -0.70, -0.40, 0.0$). Solid, dashed, and dotted lines are the case of $\log U = -1.5,~-2,~ -3$ respectively. The number densities of \textsc{Hii} region $\log n_\mathrm{HII} [\mathrm{cm^{-3}}]=1,~ 2, ~3$ are also marked by ticks from left to right on each CLOUDY line.}
   \label{fig:line_combination_JWST_ALMA}
\end{center}
\end{figure*}

It is interesting and timely to explore line-ratio diagnostics using three \OIII lines; 
88 $\mu$m, 52 \mum and 5007\AA. The former two fine-structure lines are observed by ALMA whereas the latter is to be observed by JWST. Hereafter we denote the line luminosity ratios using the wavelength such as $R_{5007/88} = L_{5007\mathrm{\mathring{A}}}/L_{88 \mu\mathrm{m}}$.
Figure \ref{fig:line_combination_JWST_ALMA} shows $R_{5007/88}$ against $R_{52/88}$ 
for our simulated galaxies.
We also show the model line ratios obtained by our set of CLOUDY calculations (Table ~\ref{table:cloudy}). 

The ratio $R_{5007/88}$ is commonly thought to be a
sensitive temperature indicator \citep[e.g.][]{Fujimoto:2022_z8p5}. Interestingly, Figure \ref{fig:line_combination_JWST_ALMA} shows that $R_{5007/88}$ may also trace the mean gas metallicity of a galaxy. We argue that it is a 
model-dependent, indirect indicator because of the complex dependence of the line emissivities on the
relevant physical quantities.
Typically, the oxygen line emissivity increases with increasing oxygen abundance (metallicity), 
but there is a critical abundance
beyond which the emissivity
{\it decreases} because of the temperature decrease of \HII regions owing to metal line cooling. The critical "peak" abundance is different for different lines
and thus line ratios vary non-trivially as
metallicity increases.

In Figure \ref{fig:line_combination_JWST_ALMA},
we plot local metal-rich galaxies observed with 
both FIR \citep{Brauher:2008} and optical emission lines \citep{Moustakas:2006}. 
Most of the plotted local galaxies have high metallicities 
with $Z > 1 Z_\odot$ and are located in the lower portion
(low $R_{52/88}$) in the figure.
Only NGC 1569, the left most symbol with $R_{5007/88} = 4.9$, has a sub-solar metallicity of $\log (Z/Z_\odot) = - 0.6$ \citep{Israel:1988}, which is located near the same metallicity line as our high-redshift galaxy samples.
The local planetary nebulae data from \citet{Dinerstein_Lester:1985} are also plotted as red stars. It can be easily estimated that the planetary nebulae have electron densities of $n_e(\OIII) \gtrsim 10^{3} \mathrm{cm^{-3}}$, which are consistent with those derived from [\textsc{Oii}] line ratios.

The line emissivities and hence the ratios 
have implicit dependence on ionization parameter through other quantities such as
electron temperature, but the dependence is weak at $\log U \sim (-3, - 2)$.
Our simulated galaxies have generally high ionization parameter with $\log U \simeq -2$ (Figure \ref{fig:FL964_OIII_compare}), and thus we may use $R_{5007/88}$ as a metallicity indicator as well.

In our emission line model (Section \ref{subsec:line_emissivity_calculation}), the \textsc{Hii} regions have a fixed density of $n_{\textsc{Hii}} = 100~\mathrm{cm^{-3}}$. Hence our galaxy samples are populated 
in the left-upper portion with $R_{52/88} \lesssim 1$.
Since $R_{52/88}$ varies weakly with $Z$ and $U$ \citep{Yang:2020}, galaxies with high $Z$ and high $U$ are distributed toward bottom/right in Figure \ref{fig:line_combination_JWST_ALMA}.


\section{Discussion}\label{sec:discussion}
In this {\it Letter}, we have studied the chemical evolution of early star-forming galaxies from $z = 9$ to $z = 6$ by using zoom-in hydrodynamics simulations. We find that oxygen line emission galaxies with stellar masses of $M_\star = 10^{9-9.5}~M_\odot$ have large ionization parameter of $\log U = -2$ and metallicity of $\log(Z/Z_\odot) \sim (-1,-0.5)$. In these galaxies, metal-enrichment occurs early and quickly over a few hundred million years. 

We have examined line diagnostics using \OIII 5007\AA, 88 $\mu$m, and 52 \mum for future observation synergies of JWST and ALMA. There have already been a few interesting observations of high-redshift galaxies.
\citet{Killi:2022} use ALMA and detect \OIII 52 \mum line from a galaxy at $z = 7$ for the first time.   
The derived value of $R_{52/88} \sim 0.7$ is close to
our galaxy samples (Figure \ref{fig:line_combination_JWST_ALMA}), and indicates a relatively low electron density of $n_e \sim 50- 260 ~{\rm cm}^{-3}$.
Observations of SMACS0723-4590 at $z=8.5$ 
by \citet{Fujimoto:2022_z8p5} 
show a large line ratio $R_{5007/88}=15.8$, which is slightly larger than our galaxy samples, suggesting a low-metallicity of $Z \sim 0.04~Z_\odot$.
Combining $R_{52/88}$ from future observation will 
constrain the values of metallicity and ionization parameter at the same time according to Figure \ref{fig:line_combination_JWST_ALMA}.
Planned observations using JWST NIRSpec are targetted to several \OIII 88 \mum emitters (e.g., GO-1657, PI: Harikane, and GO-1840, PI:\'{A}lvarez-M\'{a}rquez \& Hashimoto). Multi-line diagnostics such as those presented in this {\it Letter} holds promise to reveal the physical conditions of the ISM in the high-redshift galaxies. 

Our simulations show rapid chemical evolution at high redshift. The resulting MZR relation is consistent with up-to-date JWST observations (Figure \ref{fig:MZR}). \citet{Langan:2020} use 300 less massive galaxy samples with $M_\star \leq 10^{8.5}~M_\odot$ at $z=8$ 
and derive MZR from $z=8$ to $z=5$ (see also a similar study by \citet{Noel:2022}). 
Our galaxy samples with larger stellar masses with 
$M_\star = 10^{8.5-10.0}~M_\odot$ show a steeper MZR, which indicates rapid chemical evolution at the early epoch. 
It would be highly interesting to study relatively massive
galaxies with JWST observations such as B14–65666 \citep[]{Roberts-Borsani:2020}, A2744–YD4 \citep[]{Morishita:2022}, and MACS1149–JD1 \citep[]{Hashimoto:2018}.


There are a few caveats in our emission line model.
Most notably we do not account for dust extinction. 
Recent ALMA surveys report the existence of a substantial amount of dust in star-forming galaxies at $z\sim6-8$ \citep{Fudamoto:2020, Burgarella:2022,Bakx:2021, Schouws:2022,Tamura:2019,Inami:2022}. 
Given the importance of emission line ratios including  \OIII 5007\AA, accurate modeling of dust extinction
may be needed for future studies. 

We have studied the statistics of early emission-line galaxies and compared with recent observations.
It will be possible and important to study the internal
structure of galaxies using both JWST observations and
numerical simulations.
We have shown in Figure \ref{fig:FL964_OIII_compare} that there are large variations/fluctuations of line emissivities, metal and density distributions within a galaxy. 
\citet{Cameron:2022} argue that unresolved variations of the electron temperature within a galaxy results in a biased estimate when the so-called $T_e$-method is applied. 
JWST's NIRSpec IFU can resolve with a pixel scale of 0.1 [arcsec/pixel]\footnote{https://jwst-docs.stsci.edu/jwst-near-infrared-spectrograph}.
For our configuration shown in Figure \ref{fig:FL964_OIII_compare}, the 7.4 kpc region at $z=7$ can be resolved with $13\times 13$ pixels. 
Gravitational lensing magnification will greatly help resolving further the structure of individual galaxies. In our future work, we will generate mock two-dimensional maps for our simulated galaxies with the same resolution of NIRSpec IFU, and will address how well the physical quantities such as gas density and temperature {\it distribution} can be
reconstructed. 

\section{Acknowledgements}
We thank the referee for comments and suggestions which have improved the quality of this work. We thank Kana Moriwaki and Yuichi Harikane for fruitful discussions, and K. Nakajima who provided us the information used in Figure \ref{fig:MZR_obs}. This work made use of v2.3 of the Binary Population and Spectral Synthesis (BPASS) models as described in \citet{Byrne:2022} and \citet{Stanway:2018}. 
The authors thankfully acknowledges the computer resources at MareNostrum and the technical support provided by the Barcelona Supercomputing Center (RES-AECT-2020-3-0019).
Numerical analyses were carried out on the analysis servers at Center for Computational Astrophysics, National Astronomical Observatory of Japan.
YN acknowledges funding from JSPS KAKENHI Grant Number 23KJ0728.  
DC is a Ramon-Cajal Researcher and is supported by the Ministerio de Ciencia, Innovaci\'{o}n y Universidades (MICIU/FEDER) under research grant PID2021-122603NB-C21.

\bibliography{FL_OIII}{}

\begin{thebibliography}{}
\expandafter\ifx\csname natexlab\endcsname\relax\def\natexlab#1{#1}\fi
\providecommand{\url}[1]{\href{#1}{#1}}
\providecommand{\dodoi}[1]{doi:~\href{http://doi.org/#1}{\nolinkurl{#1}}}
\providecommand{\doeprint}[1]{\href{http://ascl.net/#1}{\nolinkurl{http://ascl.net/#1}}}
\providecommand{\doarXiv}[1]{\href{https://arxiv.org/abs/#1}{\nolinkurl{https://arxiv.org/abs/#1}}}

\bibitem[{{Anders} \& {Grevesse}(1989)}]{Andres:1989}
{Anders}, E., \& {Grevesse}, N. 1989, \gca, 53, 197,
  \dodoi{10.1016/0016-7037(89)90286-X}

\bibitem[{{Arata} {et~al.}(2020){Arata}, {Yajima}, {Nagamine}, {Abe}, \&
  {Khochfar}}]{Arata:2020}
{Arata}, S., {Yajima}, H., {Nagamine}, K., {Abe}, M., \& {Khochfar}, S. 2020,
  \mnras, 498, 5541, \dodoi{10.1093/mnras/staa2809}

\bibitem[{{Bakx} {et~al.}(2021){Bakx}, {Sommovigo}, {Carniani}, {Ferrara},
  {Akins}, {Fujimoto}, {Hagimoto}, {Knudsen}, {Pallottini}, {Tamura}, \&
  {Watson}}]{Bakx:2021}
{Bakx}, T. J.~L.~C., {Sommovigo}, L., {Carniani}, S., {et~al.} 2021, \mnras,
  508, L58, \dodoi{10.1093/mnrasl/slab104}

\bibitem[{{Barrufet} {et~al.}(2022){Barrufet}, {Oesch}, {Weibel}, {Brammer},
  {Bezanson}, {Bouwens}, {Fudamoto}, {Gonzalez}, {Illingworth}, {Heintz},
  {Holden}, {Labbe}, {Magee}, {Naidu}, {Nelson}, {Stefanon}, {Smit}, {van
  Dokkum}, {Weaver}, \& {Williams}}]{Barrufet:2022}
{Barrufet}, L., {Oesch}, P.~A., {Weibel}, A., {et~al.} 2022, arXiv e-prints,
  arXiv:2207.14733.
\newblock \doarXiv{2207.14733}

\bibitem[{{Bian} {et~al.}(2018){Bian}, {Kewley}, \& {Dopita}}]{Bian:2018}
{Bian}, F., {Kewley}, L.~J., \& {Dopita}, M.~A. 2018, \apj, 859, 175,
  \dodoi{10.3847/1538-4357/aabd74}

\bibitem[{{Brauher} {et~al.}(2008){Brauher}, {Dale}, \& {Helou}}]{Brauher:2008}
{Brauher}, J.~R., {Dale}, D.~A., \& {Helou}, G. 2008, \apjs, 178, 280,
  \dodoi{10.1086/590249}

\bibitem[{{Burgarella} {et~al.}(2022){Burgarella}, {Bogdanoska}, {Nanni},
  {Bardelli}, {B{\'e}thermin}, {Boquien}, {Buat}, {Faisst},
  {Dessauges-Zavadsky}, {Fudamoto}, {Fujimoto}, {Giavalisco}, {Ginolfi},
  {Gruppioni}, {Hathi}, {Ibar}, {Jones}, {Koekemoer}, {Kohno}, {Lemaux},
  {Narayanan}, {Oesch}, {Ouchi}, {Riechers}, {Pozzi}, {Romano}, {Schaerer},
  {Talia}, {Theul{\'e}}, {Vergani}, {Zamorani}, {Zucca}, {Cassata}, \& {ALPINE
  Team}}]{Burgarella:2022}
{Burgarella}, D., {Bogdanoska}, J., {Nanni}, A., {et~al.} 2022, \aap, 664, A73,
  \dodoi{10.1051/0004-6361/202142554}

\bibitem[{{Byrne} {et~al.}(2022){Byrne}, {Stanway}, {Eldridge}, {McSwiney}, \&
  {Townsend}}]{Byrne:2022}
{Byrne}, C.~M., {Stanway}, E.~R., {Eldridge}, J.~J., {McSwiney}, L., \&
  {Townsend}, O.~T. 2022, \mnras, 512, 5329, \dodoi{10.1093/mnras/stac807}

\bibitem[{{Calzetti} {et~al.}(2000){Calzetti}, {Armus}, {Bohlin}, {Kinney},
  {Koornneef}, \& {Storchi-Bergmann}}]{Calzetti:2000}
{Calzetti}, D., {Armus}, L., {Bohlin}, R.~C., {et~al.} 2000, \apj, 533, 682,
  \dodoi{10.1086/308692}

\bibitem[{{Cameron} {et~al.}(2022){Cameron}, {Katz}, \& {Rey}}]{Cameron:2022}
{Cameron}, A.~J., {Katz}, H., \& {Rey}, M.~P. 2022, arXiv e-prints,
  arXiv:2210.14234.
\newblock \doarXiv{2210.14234}

\bibitem[{{Carniani} {et~al.}(2017){Carniani}, {Maiolino}, {Pallottini},
  {Vallini}, {Pentericci}, {Ferrara}, {Castellano}, {Vanzella}, {Grazian},
  {Gallerani}, {Santini}, {Wagg}, \& {Fontana}}]{Carniani:2017}
{Carniani}, S., {Maiolino}, R., {Pallottini}, A., {et~al.} 2017, \aap, 605,
  A42, \dodoi{10.1051/0004-6361/201630366}

\bibitem[{{Ceverino} {et~al.}(2017){Ceverino}, {Glover}, \&
  {Klessen}}]{Ceverino:2017}
{Ceverino}, D., {Glover}, S. C.~O., \& {Klessen}, R.~S. 2017, \mnras, 470,
  2791, \dodoi{10.1093/mnras/stx1386}

\bibitem[{{Ceverino} {et~al.}(2021){Ceverino}, {Hirschmann}, {Klessen},
  {Glover}, {Charlot}, \& {Feltre}}]{Ceverino:2021}
{Ceverino}, D., {Hirschmann}, M., {Klessen}, R.~S., {et~al.} 2021, \mnras, 504,
  4472, \dodoi{10.1093/mnras/stab1206}

\bibitem[{{Ceverino} {et~al.}(2019){Ceverino}, {Klessen}, \&
  {Glover}}]{Ceverino:2019}
{Ceverino}, D., {Klessen}, R.~S., \& {Glover}, S. C.~O. 2019, \mnras, 484,
  1366, \dodoi{10.1093/mnras/stz079}

\bibitem[{{Ceverino} \& {Klypin}(2009)}]{Ceverino:2009}
{Ceverino}, D., \& {Klypin}, A. 2009, \apj, 695, 292,
  \dodoi{10.1088/0004-637X/695/1/292}

\bibitem[{{Ceverino} {et~al.}(2014){Ceverino}, {Klypin}, {Klimek},
  {Trujillo-Gomez}, {Churchill}, {Primack}, \& {Dekel}}]{Ceverino:2014}
{Ceverino}, D., {Klypin}, A., {Klimek}, E.~S., {et~al.} 2014, \mnras, 442,
  1545, \dodoi{10.1093/mnras/stu956}

\bibitem[{{Curti} {et~al.}(2020){Curti}, {Mannucci}, {Cresci}, \&
  {Maiolino}}]{Curti:2020}
{Curti}, M., {Mannucci}, F., {Cresci}, G., \& {Maiolino}, R. 2020, \mnras, 491,
  944, \dodoi{10.1093/mnras/stz2910}

\bibitem[{{Curti} {et~al.}(2022){Curti}, {D'Eugenio}, {Carniani}, {Maiolino},
  {Sandles}, {Witstok}, {Baker}, {Bennett}, {Piotrowska}, {Tacchella},
  {Charlot}, {Nakajima}, {Maheson}, {Mannucci}, {Arribas}, {Belfiore},
  {Bonaventura}, {Bunker}, {Chevallard}, {Cresci}, {Curtis-Lake},
  {Hayden-Pawson}, {Kumari}, {Laseter}, {Looser}, {Marconi}, {Maseda}, {Jones},
  {Scholtz}, {Smit}, {Ubler}, \& {Wallace}}]{Curti:2022}
{Curti}, M., {D'Eugenio}, F., {Carniani}, S., {et~al.} 2022, arXiv e-prints,
  arXiv:2207.12375.
\newblock \doarXiv{2207.12375}

\bibitem[{{De Looze} {et~al.}(2014){De Looze}, {Cormier}, {Lebouteiller},
  {Madden}, {Baes}, {Bendo}, {Boquien}, {Boselli}, {Clements}, {Cortese},
  {Cooray}, {Galametz}, {Galliano}, {Graci{\'a}-Carpio}, {Isaak}, {Karczewski},
  {Parkin}, {Pellegrini}, {R{\'e}my-Ruyer}, {Spinoglio}, {Smith}, \&
  {Sturm}}]{De_Looze:2014}
{De Looze}, I., {Cormier}, D., {Lebouteiller}, V., {et~al.} 2014, \aap, 568,
  A62, \dodoi{10.1051/0004-6361/201322489}

\bibitem[{{Dinerstein} {et~al.}(1985){Dinerstein}, {Lester}, \&
  {Werner}}]{Dinerstein_Lester:1985}
{Dinerstein}, H.~L., {Lester}, D.~F., \& {Werner}, M.~W. 1985, \apj, 291, 561,
  \dodoi{10.1086/163096}

\bibitem[{{Dopita} \& {Sutherland}(2003)}]{Dopita:2003}
{Dopita}, M.~A., \& {Sutherland}, R.~S. 2003, {Astrophysics of the diffuse
  universe}

\bibitem[{{Ferland} {et~al.}(2013){Ferland}, {Porter}, {van Hoof}, {Williams},
  {Abel}, {Lykins}, {Shaw}, {Henney}, \& {Stancil}}]{Ferland:2013}
{Ferland}, G.~J., {Porter}, R.~L., {van Hoof}, P.~A.~M., {et~al.} 2013, \rmxaa,
  49, 137.
\newblock \doarXiv{1302.4485}

\bibitem[{{Fudamoto} {et~al.}(2020){Fudamoto}, {Oesch}, {Faisst},
  {B{\'e}thermin}, {Ginolfi}, {Khusanova}, {Loiacono}, {Le F{\`e}vre}, {Capak},
  {Schaerer}, {Silverman}, {Cassata}, {Yan}, {Amorin}, {Bardelli}, {Boquien},
  {Cimatti}, {Dessauges-Zavadsky}, {Fujimoto}, {Gruppioni}, {Hathi}, {Ibar},
  {Jones}, {Koekemoer}, {Lagache}, {Lemaux}, {Maiolino}, {Narayanan}, {Pozzi},
  {Riechers}, {Rodighiero}, {Talia}, {Toft}, {Vallini}, {Vergani}, {Zamorani},
  \& {Zucca}}]{Fudamoto:2020}
{Fudamoto}, Y., {Oesch}, P.~A., {Faisst}, A., {et~al.} 2020, \aap, 643, A4,
  \dodoi{10.1051/0004-6361/202038163}

\bibitem[{{Fujimoto} {et~al.}(2022){Fujimoto}, {Ouchi}, {Nakajima}, {Harikane},
  {Isobe}, {Brammer}, {Oguri}, {Gim{\'e}nez-Arteaga}, {Heintz}, {Kokorev},
  {Bauer}, {Ferrara}, {Kojima}, {Lagos}, {Laura}, {Schaerer}, {Shimasaku},
  {Hatsukade}, {Kohno}, {Sun}, {Valentino}, {Watson}, {Fudamoto}, {Inoue},
  {Gonz{\'a}lez-L{\'o}pez}, {Koekemoer}, {Knudsen}, {Lee}, {Magdis}, {Richard},
  {Strait}, {Sugahara}, {Tamura}, {Toft}, {Umehata}, \&
  {Walth}}]{Fujimoto:2022_z8p5}
{Fujimoto}, S., {Ouchi}, M., {Nakajima}, K., {et~al.} 2022, arXiv e-prints,
  arXiv:2212.06863.
\newblock \doarXiv{2212.06863}

\bibitem[{{Graziani} {et~al.}(2020){Graziani}, {Schneider}, {Ginolfi}, {Hunt},
  {Maio}, {Glatzle}, \& {Ciardi}}]{Graziani:2020}
{Graziani}, L., {Schneider}, R., {Ginolfi}, M., {et~al.} 2020, \mnras, 494,
  1071, \dodoi{10.1093/mnras/staa796}

\bibitem[{{Gutkin} {et~al.}(2016){Gutkin}, {Charlot}, \&
  {Bruzual}}]{Gutkin:2016}
{Gutkin}, J., {Charlot}, S., \& {Bruzual}, G. 2016, \mnras, 462, 1757,
  \dodoi{10.1093/mnras/stw1716}

\bibitem[{{Harikane} {et~al.}(2020){Harikane}, {Ouchi}, {Inoue}, {Matsuoka},
  {Tamura}, {Bakx}, {Fujimoto}, {Moriwaki}, {Ono}, {Nagao}, {Tadaki}, {Kojima},
  {Shibuya}, {Egami}, {Ferrara}, {Gallerani}, {Hashimoto}, {Kohno}, {Matsuda},
  {Matsuo}, {Pallottini}, {Sugahara}, \& {Vallini}}]{Harikane:2020}
{Harikane}, Y., {Ouchi}, M., {Inoue}, A.~K., {et~al.} 2020, \apj, 896, 93,
  \dodoi{10.3847/1538-4357/ab94bd}

\bibitem[{{Harikane} {et~al.}(2022){Harikane}, {Inoue}, {Mawatari},
  {Hashimoto}, {Yamanaka}, {Fudamoto}, {Matsuo}, {Tamura}, {Dayal}, {Yung},
  {Hutter}, {Pacucci}, {Sugahara}, \& {Koekemoer}}]{Harikane:2022_HDrop}
{Harikane}, Y., {Inoue}, A.~K., {Mawatari}, K., {et~al.} 2022, \apj, 929, 1,
  \dodoi{10.3847/1538-4357/ac53a9}

\bibitem[{{Hashimoto} {et~al.}(2018){Hashimoto}, {Laporte}, {Mawatari},
  {Ellis}, {Inoue}, {Zackrisson}, {Roberts-Borsani}, {Zheng}, {Tamura},
  {Bauer}, {Fletcher}, {Harikane}, {Hatsukade}, {Hayatsu}, {Matsuda}, {Matsuo},
  {Okamoto}, {Ouchi}, {Pell{\'o}}, {Rydberg}, {Shimizu}, {Taniguchi},
  {Umehata}, \& {Yoshida}}]{Hashimoto:2018}
{Hashimoto}, T., {Laporte}, N., {Mawatari}, K., {et~al.} 2018, \nat, 557, 392,
  \dodoi{10.1038/s41586-018-0117-z}

\bibitem[{{Hashimoto} {et~al.}(2019){Hashimoto}, {Inoue}, {Mawatari}, {Tamura},
  {Matsuo}, {Furusawa}, {Harikane}, {Shibuya}, {Knudsen}, {Kohno}, {Ono},
  {Zackrisson}, {Okamoto}, {Kashikawa}, {Oesch}, {Ouchi}, {Ota}, {Shimizu},
  {Taniguchi}, {Umehata}, \& {Watson}}]{Hashimoto:2019}
{Hashimoto}, T., {Inoue}, A.~K., {Mawatari}, K., {et~al.} 2019, \pasj, 71, 71,
  \dodoi{10.1093/pasj/psz049}

\bibitem[{{Heintz} {et~al.}(2022){Heintz}, {Gim{\'e}nez-Arteaga}, {Fujimoto},
  {Brammer}, {Espada}, {Gillman}, {Gonz{\'a}lez-L{\'o}pez}, {Greve},
  {Harikane}, {Hatsukade}, {Knudsen}, {Koekemoer}, {Kohno}, {Kokorev}, {Lee},
  {Magdis}, {Nelson}, {Rizzo}, {Sanders}, {Schaerer}, {Shapley}, {Strait},
  {Sun}, {Toft}, {Valentino}, {Vijayan}, {Watson}, {Bauer}, {Christiansen}, \&
  {Wilson}}]{Heintz:2022}
{Heintz}, K.~E., {Gim{\'e}nez-Arteaga}, C., {Fujimoto}, S., {et~al.} 2022,
  arXiv e-prints, arXiv:2212.06877.
\newblock \doarXiv{2212.06877}

\bibitem[{{Hirschmann} {et~al.}(2017){Hirschmann}, {Charlot}, {Feltre}, {Naab},
  {Choi}, {Ostriker}, \& {Somerville}}]{Hirschmann:2017}
{Hirschmann}, M., {Charlot}, S., {Feltre}, A., {et~al.} 2017, \mnras, 472,
  2468, \dodoi{10.1093/mnras/stx2180}

\bibitem[{{Hirschmann} {et~al.}(2022){Hirschmann}, {Charlot}, {Feltre},
  {Curtis-Lake}, {Somerville}, {Chevallard}, {Choi}, {Nelson}, {Morisset},
  {Plat}, \& {Vidal-Garcia}}]{Hirschmann:2022}
---. 2022, arXiv e-prints, arXiv:2212.02522.
\newblock \doarXiv{2212.02522}

\bibitem[{{Inami} {et~al.}(2022){Inami}, {Algera}, {Schouws}, {Sommovigo},
  {Bouwens}, {Smit}, {Stefanon}, {Bowler}, {Endsley}, {Ferrara}, {Oesch},
  {Stark}, {Aravena}, {Barrufet}, {da Cunha}, {Dayal}, {De Looze}, {Fudamoto},
  {Gonzalez}, {Graziani}, {Hodge}, {Hygate}, {Nanayakkara}, {Pallottini},
  {Riechers}, {Schneider}, {Topping}, \& {van der Werf}}]{Inami:2022}
{Inami}, H., {Algera}, H. S.~B., {Schouws}, S., {et~al.} 2022, \mnras, 515,
  3126, \dodoi{10.1093/mnras/stac1779}

\bibitem[{{Inoue} {et~al.}(2014){Inoue}, {Shimizu}, {Tamura}, {Matsuo},
  {Okamoto}, \& {Yoshida}}]{Inoue:2014}
{Inoue}, A.~K., {Shimizu}, I., {Tamura}, Y., {et~al.} 2014, \apjl, 780, L18,
  \dodoi{10.1088/2041-8205/780/2/L18}

\bibitem[{{Inoue} {et~al.}(2016){Inoue}, {Tamura}, {Matsuo}, {Mawatari},
  {Shimizu}, {Shibuya}, {Ota}, {Yoshida}, {Zackrisson}, {Kashikawa}, {Kohno},
  {Umehata}, {Hatsukade}, {Iye}, {Matsuda}, {Okamoto}, \&
  {Yamaguchi}}]{Inoue:2016}
{Inoue}, A.~K., {Tamura}, Y., {Matsuo}, H., {et~al.} 2016, Science, 352, 1559,
  \dodoi{10.1126/science.aaf0714}

\bibitem[{{Israel}(1988)}]{Israel:1988}
{Israel}, F.~P. 1988, \aap, 194, 24

\bibitem[{{Izotov} {et~al.}(2019){Izotov}, {Guseva}, {Fricke}, \&
  {Henkel}}]{Izotov:2019}
{Izotov}, Y.~I., {Guseva}, N.~G., {Fricke}, K.~J., \& {Henkel}, C. 2019, \aap,
  623, A40, \dodoi{10.1051/0004-6361/201834768}

\bibitem[{{Katz} {et~al.}(2019){Katz}, {Galligan}, {Kimm}, {Rosdahl},
  {Haehnelt}, {Blaizot}, {Devriendt}, {Slyz}, {Laporte}, \&
  {Ellis}}]{Katz:2019}
{Katz}, H., {Galligan}, T.~P., {Kimm}, T., {et~al.} 2019, \mnras, 487, 5902,
  \dodoi{10.1093/mnras/stz1672}

\bibitem[{{Killi} {et~al.}(2022){Killi}, {Watson}, {Fujimoto}, {Akins},
  {Knudsen}, {Richard}, {Harikane}, {Rigopoulou}, {Rizzo}, {Ginolfi},
  {Popping}, \& {Kokorev}}]{Killi:2022}
{Killi}, M., {Watson}, D., {Fujimoto}, S., {et~al.} 2022, arXiv e-prints,
  arXiv:2211.01424.
\newblock \doarXiv{2211.01424}

\bibitem[{{Kimm} \& {Cen}(2014)}]{Kimm_Cen:2014}
{Kimm}, T., \& {Cen}, R. 2014, \apj, 788, 121,
  \dodoi{10.1088/0004-637X/788/2/121}

\bibitem[{{Klypin} {et~al.}(2011){Klypin}, {Trujillo-Gomez}, \&
  {Primack}}]{Klypin:2011}
{Klypin}, A.~A., {Trujillo-Gomez}, S., \& {Primack}, J. 2011, \apj, 740, 102,
  \dodoi{10.1088/0004-637X/740/2/102}

\bibitem[{{Kohandel} {et~al.}(2022){Kohandel}, {Ferrara}, {Pallottini},
  {Vallini}, {Sommovigo}, \& {Ziparo}}]{Kohandel:2022}
{Kohandel}, M., {Ferrara}, A., {Pallottini}, A., {et~al.} 2022, arXiv e-prints,
  arXiv:2212.02519.
\newblock \doarXiv{2212.02519}

\bibitem[{{Kravtsov}(2003)}]{Kravtsov:2003}
{Kravtsov}, A.~V. 2003, \apjl, 590, L1, \dodoi{10.1086/376674}

\bibitem[{{Kravtsov} {et~al.}(1997){Kravtsov}, {Klypin}, \&
  {Khokhlov}}]{Kravtsov:1997}
{Kravtsov}, A.~V., {Klypin}, A.~A., \& {Khokhlov}, A.~M. 1997, \apjs, 111, 73,
  \dodoi{10.1086/313015}

\bibitem[{{Langan} {et~al.}(2020){Langan}, {Ceverino}, \&
  {Finlator}}]{Langan:2020}
{Langan}, I., {Ceverino}, D., \& {Finlator}, K. 2020, \mnras, 494, 1988,
  \dodoi{10.1093/mnras/staa880}

\bibitem[{{Langeroodi} {et~al.}(2022){Langeroodi}, {Hjorth}, {Chen}, {Kelly},
  {Williams}, {Lin}, {Scarlata}, {Zitrin}, {Broadhurst}, {Diego}, {Huang},
  {Filippenko}, {Foley}, {Jha}, {Koekemoer}, {Oguri}, {Perez-Fournon},
  {Pierel}, {Poidevin}, \& {Strolger}}]{Langeroodi:2022}
{Langeroodi}, D., {Hjorth}, J., {Chen}, W., {et~al.} 2022, arXiv e-prints,
  arXiv:2212.02491.
\newblock \doarXiv{2212.02491}

\bibitem[{{Laporte} {et~al.}(2017){Laporte}, {Ellis}, {Boone}, {Bauer},
  {Qu{\'e}nard}, {Roberts-Borsani}, {Pell{\'o}}, {P{\'e}rez-Fournon}, \&
  {Streblyanska}}]{Laporte:2017}
{Laporte}, N., {Ellis}, R.~S., {Boone}, F., {et~al.} 2017, \apjl, 837, L21,
  \dodoi{10.3847/2041-8213/aa62aa}

\bibitem[{{Leethochawalit} {et~al.}(2022){Leethochawalit}, {Trenti}, {Santini},
  {Yang}, {Merlin}, {Castellano}, {Fontana}, {Treu}, {Mason}, {Glazebrook},
  {Jones}, {Vulcani}, {Nanayakkara}, {Marchesini}, {Mascia}, {Morishita},
  {Roberts-Borsani}, {Bonchi}, {Paris}, {Boyett}, {Strait}, {Calabro`},
  {Pentericci}, {Bradac}, {Wang}, \& {Scarlata}}]{Leethochawaliit:2022}
{Leethochawalit}, N., {Trenti}, M., {Santini}, P., {et~al.} 2022, arXiv
  e-prints, arXiv:2207.11135.
\newblock \doarXiv{2207.11135}

\bibitem[{{Mandelker} {et~al.}(2017){Mandelker}, {Dekel}, {Ceverino}, {DeGraf},
  {Guo}, \& {Primack}}]{Mandelker:2017}
{Mandelker}, N., {Dekel}, A., {Ceverino}, D., {et~al.} 2017, \mnras, 464, 635,
  \dodoi{10.1093/mnras/stw2358}

\bibitem[{{Mandelker} {et~al.}(2014){Mandelker}, {Dekel}, {Ceverino}, {Tweed},
  {Moody}, \& {Primack}}]{Mandelker:2014}
---. 2014, \mnras, 443, 3675, \dodoi{10.1093/mnras/stu1340}

\bibitem[{{Morishita} {et~al.}(2022){Morishita}, {Roberts-Borsani}, {Treu},
  {Brammer}, {Mason}, {Trenti}, {Vulcani}, {Wang}, {Acebron}, {Bah{\'e}},
  {Bergamini}, {Boyett}, {Bradac}, {Calabr{\`o}}, {Castellano}, {Chen}, {De
  Lucia}, {Filippenko}, {Fontana}, {Glazebrook}, {Grillo}, {Henry}, {Jones},
  {Kelly}, {Koekemoer}, {Leethochawalit}, {Lu}, {Marchesini}, {Mascia},
  {Mercurio}, {Merlin}, {Metha}, {Nanayakkara}, {Nonino}, {Paris},
  {Pentericci}, {Santini}, {Strait}, {Vanzella}, {Windhorst}, {Rosati}, \&
  {Xie}}]{Morishita:2022}
{Morishita}, T., {Roberts-Borsani}, G., {Treu}, T., {et~al.} 2022, arXiv
  e-prints, arXiv:2211.09097.
\newblock \doarXiv{2211.09097}

\bibitem[{{Moriwaki} {et~al.}(2018){Moriwaki}, {Yoshida}, {Shimizu},
  {Harikane}, {Matsuda}, {Matsuo}, {Hashimoto}, {Inoue}, {Tamura}, \&
  {Nagao}}]{Moriwaki:2018}
{Moriwaki}, K., {Yoshida}, N., {Shimizu}, I., {et~al.} 2018, \mnras, 481, L84,
  \dodoi{10.1093/mnrasl/sly167}

\bibitem[{{Moustakas} {et~al.}(2006){Moustakas}, {Kennicutt}, \&
  {Tremonti}}]{Moustakas:2006}
{Moustakas}, J., {Kennicutt}, Robert~C., J., \& {Tremonti}, C.~A. 2006, \apj,
  642, 775, \dodoi{10.1086/500964}

\bibitem[{{Mushtaq} {et~al.}(2023){Mushtaq}, {Ceverino}, {Klessen}, {Reissl},
  \& {Puttasiddappa}}]{Mushtaq:2023}
{Mushtaq}, M., {Ceverino}, D., {Klessen}, R.~S., {Reissl}, S., \&
  {Puttasiddappa}, P.~H. 2023, arXiv e-prints, arXiv:2304.10150,
  \dodoi{10.48550/arXiv.2304.10150}

\bibitem[{{Nakajima} {et~al.}(2020){Nakajima}, {Ellis}, {Robertson}, {Tang}, \&
  {Stark}}]{Nakajima:2020}
{Nakajima}, K., {Ellis}, R.~S., {Robertson}, B.~E., {Tang}, M., \& {Stark},
  D.~P. 2020, \apj, 889, 161, \dodoi{10.3847/1538-4357/ab6604}

\bibitem[{{Nakajima} {et~al.}(2023){Nakajima}, {Ouchi}, {Isobe}, {Harikane},
  {Zhang}, {Ono}, {Umeda}, \& {Oguri}}]{Nakajima:2023}
{Nakajima}, K., {Ouchi}, M., {Isobe}, Y., {et~al.} 2023, arXiv e-prints,
  arXiv:2301.12825, \dodoi{10.48550/arXiv.2301.12825}

\bibitem[{{Noel} {et~al.}(2022){Noel}, {Zhu}, \& {Gnedin}}]{Noel:2022}
{Noel}, I., {Zhu}, H., \& {Gnedin}, N. 2022, arXiv e-prints, arXiv:2210.16750.
\newblock \doarXiv{2210.16750}

\bibitem[{{Nomoto} {et~al.}(2006){Nomoto}, {Tominaga}, {Umeda}, {Kobayashi}, \&
  {Maeda}}]{Nomoto:2006}
{Nomoto}, K., {Tominaga}, N., {Umeda}, H., {Kobayashi}, C., \& {Maeda}, K.
  2006, \nphysa, 777, 424, \dodoi{10.1016/j.nuclphysa.2006.05.008}

\bibitem[{{Olsen} {et~al.}(2017){Olsen}, {Greve}, {Narayanan}, {Thompson},
  {Dav{\'e}}, {Niebla Rios}, \& {Stawinski}}]{Olsen:2017}
{Olsen}, K., {Greve}, T.~R., {Narayanan}, D., {et~al.} 2017, \apj, 846, 105,
  \dodoi{10.3847/1538-4357/aa86b4}

\bibitem[{{Osterbrock} \& {Ferland}(2006)}]{Osterbrock:2006}
{Osterbrock}, D.~E., \& {Ferland}, G.~J. 2006, {Astrophysics of gaseous nebulae
  and active galactic nuclei}

\bibitem[{{Paardekooper} {et~al.}(2015){Paardekooper}, {Khochfar}, \& {Dalla
  Vecchia}}]{Paardekooper:2015}
{Paardekooper}, J.-P., {Khochfar}, S., \& {Dalla Vecchia}, C. 2015, \mnras,
  451, 2544, \dodoi{10.1093/mnras/stv1114}

\bibitem[{{Pallottini} {et~al.}(2022){Pallottini}, {Ferrara}, {Gallerani},
  {Behrens}, {Kohandel}, {Carniani}, {Vallini}, {Salvadori}, {Gelli},
  {Sommovigo}, {D'Odorico}, {Di Mascia}, \& {Pizzati}}]{Pallottini:2022}
{Pallottini}, A., {Ferrara}, A., {Gallerani}, S., {et~al.} 2022, \mnras, 513,
  5621, \dodoi{10.1093/mnras/stac1281}

\bibitem[{{Panuzzo} {et~al.}(2003){Panuzzo}, {Bressan}, {Granato}, {Silva}, \&
  {Danese}}]{Panuzzo:2003}
{Panuzzo}, P., {Bressan}, A., {Granato}, G.~L., {Silva}, L., \& {Danese}, L.
  2003, \aap, 409, 99, \dodoi{10.1051/0004-6361:20031094}

\bibitem[{{Roberts-Borsani} {et~al.}(2020){Roberts-Borsani}, {Ellis}, \&
  {Laporte}}]{Roberts-Borsani:2020}
{Roberts-Borsani}, G.~W., {Ellis}, R.~S., \& {Laporte}, N. 2020, \mnras, 497,
  3440, \dodoi{10.1093/mnras/staa2085}

\bibitem[{{Schaerer} {et~al.}(2022){Schaerer}, {Marques-Chaves}, {Oesch},
  {Naidu}, {Barrufet}, {Izotov}, {Guseva}, \& {Brammer}}]{Schaerer:2022}
{Schaerer}, D., {Marques-Chaves}, R., {Oesch}, P., {et~al.} 2022, arXiv
  e-prints, arXiv:2207.10034.
\newblock \doarXiv{2207.10034}

\bibitem[{{Schouws} {et~al.}(2022){Schouws}, {Stefanon}, {Bouwens}, {Smit},
  {Hodge}, {Labb{\'e}}, {Algera}, {Boogaard}, {Carniani}, {Fudamoto},
  {Holwerda}, {Illingworth}, {Maiolino}, {Maseda}, {Oesch}, \& {van der
  Werf}}]{Schouws:2022}
{Schouws}, S., {Stefanon}, M., {Bouwens}, R., {et~al.} 2022, \apj, 928, 31,
  \dodoi{10.3847/1538-4357/ac4605}

\bibitem[{{Shimizu} {et~al.}(2016){Shimizu}, {Inoue}, {Okamoto}, \&
  {Yoshida}}]{Shimizu:2016}
{Shimizu}, I., {Inoue}, A.~K., {Okamoto}, T., \& {Yoshida}, N. 2016, \mnras,
  461, 3563, \dodoi{10.1093/mnras/stw1423}

\bibitem[{{Stanway} \& {Eldridge}(2018)}]{Stanway:2018}
{Stanway}, E.~R., \& {Eldridge}, J.~J. 2018, \mnras, 479, 75,
  \dodoi{10.1093/mnras/sty1353}

\bibitem[{{Sugahara} {et~al.}(2022){Sugahara}, {Inoue}, {Fudamoto},
  {Hashimoto}, {Harikane}, \& {Yamanaka}}]{Sugahara:2022}
{Sugahara}, Y., {Inoue}, A.~K., {Fudamoto}, Y., {et~al.} 2022, \apj, 935, 119,
  \dodoi{10.3847/1538-4357/ac7fed}

\bibitem[{{Sun} {et~al.}(2022){Sun}, {Egami}, {Pirzkal}, {Rieke}, {Baum},
  {Boyer}, {Boyett}, {Bunker}, {Cameron}, {Curti}, {Eisenstein}, {Gennaro},
  {Greene}, {Jaffe}, {Kelly}, {Koekemoer}, {Kumari}, {Maiolino}, {Maseda},
  {Perna}, {Rest}, {Robertson}, {Schlawin}, {Smit}, {Stansberry}, {Sunnquist},
  {Tacchella}, {Williams}, \& {Willmer}}]{Sun:2022b}
{Sun}, F., {Egami}, E., {Pirzkal}, N., {et~al.} 2022, arXiv e-prints,
  arXiv:2209.03374.
\newblock \doarXiv{2209.03374}

\bibitem[{{Tacchella} {et~al.}(2022){Tacchella}, {Finkelstein}, {Bagley},
  {Dickinson}, {Ferguson}, {Giavalisco}, {Graziani}, {Grogin}, {Hathi},
  {Hutchison}, {Jung}, {Koekemoer}, {Larson}, {Papovich}, {Pirzkal},
  {Rojas-Ruiz}, {Song}, {Schneider}, {Somerville}, {Wilkins}, \&
  {Yung}}]{Tacchella:2022}
{Tacchella}, S., {Finkelstein}, S.~L., {Bagley}, M., {et~al.} 2022, \apj, 927,
  170, \dodoi{10.3847/1538-4357/ac4cad}

\bibitem[{{Tamura} {et~al.}(2019){Tamura}, {Mawatari}, {Hashimoto}, {Inoue},
  {Zackrisson}, {Christensen}, {Binggeli}, {Matsuda}, {Matsuo}, {Takeuchi},
  {Asano}, {Sunaga}, {Shimizu}, {Okamoto}, {Yoshida}, {Lee}, {Shibuya},
  {Taniguchi}, {Umehata}, {Hatsukade}, {Kohno}, \& {Ota}}]{Tamura:2019}
{Tamura}, Y., {Mawatari}, K., {Hashimoto}, T., {et~al.} 2019, \apj, 874, 27,
  \dodoi{10.3847/1538-4357/ab0374}

\bibitem[{{Thielemann} {et~al.}(1986){Thielemann}, {Nomoto}, \&
  {Yokoi}}]{Thielemann:1986}
{Thielemann}, F.~K., {Nomoto}, K., \& {Yokoi}, K. 1986, \aap, 158, 17

\bibitem[{{Topping} {et~al.}(2022){Topping}, {Stark}, {Endsley}, {Bouwens},
  {Schouws}, {Smit}, {Stefanon}, {Inami}, {Bowler}, {Oesch}, {Gonzalez},
  {Dayal}, {da Cunha}, {Algera}, {van der Werf}, {Pallottini}, {Barrufet},
  {Schneider}, {De Looze}, {Sommovigo}, {Whitler}, {Graziani}, {Fudamoto}, \&
  {Ferrara}}]{Topping:2022}
{Topping}, M.~W., {Stark}, D.~P., {Endsley}, R., {et~al.} 2022, \mnras, 516,
  975, \dodoi{10.1093/mnras/stac2291}

\bibitem[{{Trussler} {et~al.}(2022){Trussler}, {Adams}, {Conselice},
  {Ferreira}, {Austin}, {Bhatawdekar}, {Caruana}, {Lovell}, {Roper}, {Verma},
  {Vijayan}, \& {Wilkins}}]{Trussler:2022}
{Trussler}, J. A.~A., {Adams}, N.~J., {Conselice}, C.~J., {et~al.} 2022, arXiv
  e-prints, arXiv:2207.14265.
\newblock \doarXiv{2207.14265}

\bibitem[{{Williams} {et~al.}(2022){Williams}, {Kelly}, {Chen}, {Brammer},
  {Zitrin}, {Treu}, {Scarlata}, {Koekemoer}, {Oguri}, {Lin}, {Diego}, {Nonino},
  {Hjorth}, {Broadhurst}, {Rogers}, {Perez-Fournon}, {Foley}, {Jha},
  {Filippenko}, {Strolger}, {Pierel}, \& {Poidevin}}]{Williams:2022}
{Williams}, H., {Kelly}, P.~L., {Chen}, W., {et~al.} 2022, arXiv e-prints,
  arXiv:2210.15699.
\newblock \doarXiv{2210.15699}

\bibitem[{{Wise} {et~al.}(2014){Wise}, {Demchenko}, {Halicek}, {Norman},
  {Turk}, {Abel}, \& {Smith}}]{Wise:2014}
{Wise}, J.~H., {Demchenko}, V.~G., {Halicek}, M.~T., {et~al.} 2014, \mnras,
  442, 2560, \dodoi{10.1093/mnras/stu979}

\bibitem[{{Witstok} {et~al.}(2022){Witstok}, {Smit}, {Maiolino}, {Kumari},
  {Aravena}, {Boogaard}, {Bouwens}, {Carniani}, {Hodge}, {Jones}, {Stefanon},
  {van der Werf}, \& {Schouws}}]{Witstok:2022}
{Witstok}, J., {Smit}, R., {Maiolino}, R., {et~al.} 2022, \mnras,
  \dodoi{10.1093/mnras/stac1905}

\bibitem[{{Wong} {et~al.}(2022){Wong}, {Wang}, {Hashimoto}, {Takagi}, {Goto},
  {Kim}, {Wu}, {On}, {Santos}, {Lu}, {Kilerci-Eser}, {Ho}, \&
  {Hsiao}}]{Wong:2022}
{Wong}, Y. H.~V., {Wang}, P., {Hashimoto}, T., {et~al.} 2022, \apj, 929, 161,
  \dodoi{10.3847/1538-4357/ac5cc7}

\bibitem[{{Woosley} \& {Weaver}(1995)}]{Woosley:1995}
{Woosley}, S.~E., \& {Weaver}, T.~A. 1995, \apjs, 101, 181,
  \dodoi{10.1086/192237}

\bibitem[{{Xiao} {et~al.}(2018){Xiao}, {Stanway}, \& {Eldridge}}]{Xiao:2018}
{Xiao}, L., {Stanway}, E.~R., \& {Eldridge}, J.~J. 2018, \mnras, 477, 904,
  \dodoi{10.1093/mnras/sty646}

\bibitem[{{Xu} {et~al.}(2016){Xu}, {Wise}, {Norman}, {Ahn}, \&
  {O'Shea}}]{Xu:2016}
{Xu}, H., {Wise}, J.~H., {Norman}, M.~L., {Ahn}, K., \& {O'Shea}, B.~W. 2016,
  \apj, 833, 84, \dodoi{10.3847/1538-4357/833/1/84}

\bibitem[{{Yajima} {et~al.}(2011){Yajima}, {Choi}, \& {Nagamine}}]{Yajima:2011}
{Yajima}, H., {Choi}, J.-H., \& {Nagamine}, K. 2011, \mnras, 412, 411,
  \dodoi{10.1111/j.1365-2966.2010.17920.x}

\bibitem[{{Yang} \& {Lidz}(2020)}]{Yang:2020}
{Yang}, S., \& {Lidz}, A. 2020, \mnras, 499, 3417,
  \dodoi{10.1093/mnras/staa3000}

\end{thebibliography}
\bibliographystyle{aasjournal}

\end{document}